\documentclass[aps,preprintnumbers,
amsmath,amssymb,prd,superscriptaddress,longbibliography
]{revtex4-2}
\UseRawInputEncoding
\usepackage{dcolumn}
\usepackage{color}

\usepackage[caption=false]{subfig}
\usepackage{feynmp}
\usepackage{feynmp-auto}
\usepackage{float}
\usepackage{bbm}
\usepackage{empheq}
\usepackage{appendix}
\def\comment#1{}

\newcommand{\nc}{\newcommand}
\nc{\scs}{\scriptstyle}
\nc{\setval}{\fmfset{wiggly_len}{3mm} \fmfset{arrow_len}{1.5mm}
	\fmfset{arrow_ang}{13} \fmfset{dash_len}{1.5mm}\fmfpen{0.125mm}
	\fmfset{dot_size}{2thick}}

\usepackage{bm,latexsym,mathrsfs,enumerate,color}
\usepackage[mathcal]{euscript}
\usepackage[breaklinks=true,unicode=true,urlcolor = blue,colorlinks = true,citecolor = blue,linkcolor = blue]{hyperref}
\usepackage{graphicx}
\usepackage{todonotes}
\usepackage{wrapfig}

\renewcommand{\vec}[1]{\bm{#1}}

\def\slashchar#1{\setbox0=\hbox{$#1$}           
	\dimen0=\wd0                                 
	\setbox1=\hbox{/} \dimen1=\wd1               
	\ifdim\dimen0>\dimen1                        
	\rlap{\hbox to \dimen0{\hfil/\hfil}}      
	#1                                        
	\else                                        
	\rlap{\hbox to \dimen1{\hfil$#1$\hfil}}   
	/                                         
	\fi}                                         %

\DeclareMathAlphabet\mathbfcal{OMS}{cmsy}{b}{n}

\usepackage{physics}
\usepackage{xcolor}

\usepackage{physics}

\begin{document}

\title{ Skyrmion-vortex hybrid and spin wave solutions in ferromagnetic superconductors}
\author{Shantonu Mukherjee}
\email{shantanumukherjeephy@gmail.com}
\affiliation{S N Bose National Centre for Basic Sciences, Block JD, Sector III, Salt Lake, Kolkata 700106, India}
\affiliation{Institute of Theoretical Solid State Physics, IFW Dresden, Helmholtzstra\ss e 20, 01069 Dresden, Germany}
\affiliation{Institute for Solid State and Materials Physics, TU Dresden D-01062 Dresden, Germany}

\author{Amitabha Lahiri}
\email{amitabha@bose.res.in}
\affiliation{S N Bose National Centre for Basic Sciences, Block JD, Sector III, Salt Lake, Kolkata 700106, India}

\begin{abstract}
    The coexistence of Ferromagnetism and superconductivity in so called ferromagnetic superconductors (FMSC) is an intriguing phenomenon which may lead to novel physical effects as well as applications. Here in this work we have explored the interplay of topological excitations, namely vortices and Skyrmions, in ferromagnetic superconductors using a field theoretic description of such systems. In particular, numerical solutions for the continuous spin field compatible to a given vortex profile are determined in absence and presence of a Dzyaloshinskii-Moriya interaction (DMI) term. The solutions show that the spin configuration is like a Skyrmion but intertwined with the vortex structure --- the radius of the the Skyrmion-like solution depends on the penetration depth and also the polarity of the Skyrmion depends on the sign of the winding number. Thus our solution describes a Skyrmion-vortex composite --- but as a topological object in a bulk ferromagnetic superconductor, rather than on the interface of a ferromagnet-superconductor heterostructure.    
    We have also determined the spin wave solutions in such systems in presence and absence of a vortex. In absence of a vortex, the frequency and the wave vector satisfy a cubic equation which leads to various interesting features. In particular, we have shown that in the low frequency regime the minimum in dispersion relation shifts from $k=0$ to a non-zero $k$ value depending on the parameters. We also discuss the nature of spin wave dispersion when the frequency $\omega$ is close to the photon mass $\tilde{m}$\,, for which we find multiple local minima in the dispersion curve. The group velocity of the spin wave would change its sign across such a minimum which is unique to FMSC. Also, the spin wave modes around the local minimum looks like roton mode in superfluid and hence called a magnetic roton. In presence of a vortex, the spin wave amplitude is shown to vary spatially such that the profile looks like that of a N\'eel Skyrmion. Possible experimental signature of both solutions are also discussed.
\end{abstract}

\date{\today}

\maketitle

\section{Introduction}
 
 Superconductivity and ferromagnetism are among the most important phenomena encountered in condensed matter physics. Apriori, the properties of superconductors and ferromagnets seem to be totally incompatible with each other. The diamagnetic property of a superconductor prevents magnetic fields from penetrating into the bulk; on the other hand in a ferromagnet, local spins arrange themselves in a certain orientation producing a net magnetization which in turn can produce a magnetic field of their own. Thus it can be expected that controlled interaction among these phenomena may lead to new physics, as well as opening new ways for technological applications. One of the possible ways to explore such an interaction is by forming a heterostructure consisting of ferromagnetic and  superconducting layers coupled via an electromagnetic coupling ~\cite{PhysRevLett.122.097001} or/and by means of a proximity coupling~\cite{PhysRevB.99.014511, PhysRevLett.117.017001}, which is expressed by the Lifshitz invariant~\cite{ 10.1088/0953-8984/8/3/012}. Hybrids of superconductors and ferromagnets have received much attention in the recent past due to their importance in exploring their fundamental aspects, as well as in applications attempting a realization of Majorana fermions~\cite{ PhysRevB.93.224505, PhysRevB.97.115136, PhysRevLett.111.206802,PhysRevResearch.5.033109}, in superconductor  spintronics~\cite{2015NatPh..11..307L, 2006Natur.439..825K,PhysRevB.72.092508}, etc. A number of papers in this direction\cite{PhysRevB.100.014431,PhysRevB.103.174519,PhysRevB.102.014503,PhysRevLett.126.117205,PhysRevApplied.17.034069} have considered the appearance of Skyrmions and vortices in ferromagnetic and superconducting layers respectively and studied their interaction. In particular, the possibility of formation of a composite of these topological excitations was studied~\cite{PhysRevLett.122.097001,PhysRevLett.117.017001,PhysRevB.100.014431,PhysRevLett.126.117205}. Composite objects formed out of vortices and Skyrmion are fundamentally interesting and may lead to intriguing applications.\\

The other way of exploring this interplay of ferromagnetism and superconductivity is to find systems which have a phase where these properties coexist. Such systems are called ferromagnetic
superconductors -- these are very difficult to find and stabilize in the mentioned phase. One of the possible 
ways to realize this phase is to use spin triplet superconductivity where the Cooper pairs have a net spin angular momentum $S=1$. These Cooper pairs, having a net spin magnetic moment can give rise to ferromagnetism as well as long range superconducting order~\cite{1998Natur.396..658I, PhysRevLett.104.137002}. 
There also exist a number of material systems where the electrons which are responsible for superconductivity are different from the static unpaired electrons which lead to ferromagnetism. This mechanism allows the conventional superconducting order to coexist with ferromagnetic ordering. Examples include Rare Earth Rhodium Boride (ErRh$_4$B$_4$) and some other rare earth materials (HoMo$_6$S$_4$, HoMo$_6$Se$_8$)~\cite{ISHIKAWA197737, SHELTON1976213, PhysRevLett.38.987,PhysRevLett.38.987, Vagov2023284}. In such materials superconductivity exists between an upper critical temperature ($T_{c1}$) and a lower critical temperature ($T_{c2}$) and is called reentrant superconductivity. Below $T_{c2}$ a ferromagnetic phase appears and near to this temperature long range ferromagnetic ordering exist with superconductivity in a narrow range of temperature ($\Delta T \sim 50$ mK). More recently, a number of materials, prepared as thin films, is shown to have similar property of coexisting ferromagnetism and superconductivity~\cite{Oner_2020, Zhu2016SignatureOC, PhysRevB.74.094518}. One of the examples is the ultra-thin samples of NbSe$_2$ with structural modulation via adsorption of Hydrazine molecules~\cite{Zhu2016SignatureOC}. \\

In this work we are interested in the second kind of ferromagnetic superconductors (FMSC), in which the electrons responsible for superconducting order are different from those responsible for ferromagnetic order, for studying the interplay of superconductivity and magnetism. Although a heterostructure of superconductor and ferromagnet is convenient in terms of formation and control, similar studies in a bulk ferromagnetic superconductor can reveal rich physics as the ferromagnetic and superconducting orders are more intricately coupled. One intriguing example of this intricate coupling came from the works by Varma and others \cite{PhysRevLett.46.49, PhysRevB.58.11624} where they have discussed a phase called the spontaneous vortex phase (SVP) in which vortices are generated self-consistently by the underlying magnetization. Existence of this SVP phase motivates us to investigate the emergence of a unique topological configuration resulting from the coupling between superconductivity and magnetism within a field-theoretic framework. This configuration may manifest as a fusion of a vortex and a Skyrmion. It may also be possible that this configuration could host local Majorana modes in ferromagnetic superconductor bulk as is known to happen on the interface~\cite{PhysRevB.85.144505}.  Previous studies have examined interactions between vortices and Skyrmions in heterostructures, where these topological objects form in separate layers. For instance, Baumard et al.~\cite{PhysRevB.99.014511} studied vortex generation by N\'eel Skyrmions, while Dahir et al.\cite{PhysRevLett.122.097001} and Andriyakhina et al.~\cite{PhysRevB.103.174519} explored Skyrmion -- Pearl vortex composites involving Bloch and N\'eel Skyrmions, respectively. In contrast, our focus lies on a three-dimensional bulk system where superconductivity and ferromagnetism coexist. We aim to derive a vortex-Skyrmion composite structure as a direct outcome of the field equations. Field-theoretic approaches are particularly well-suited for capturing the universal low-energy features of condensed matter systems. In our case, this framework not only facilitates the derivation of novel composite solutions but also enables an analysis of their stability using tools like Derrick's theorem.\\

Here in this work we shall model the FMSC using a field theoretic framework which takes into account the low energy fluctuations of ferromagnetic and superconducting order. It is known for a long time that magnetic Skyrmions naturally emerge as topological configurations in the nonlinear sigma model. In presence of a Dzyaloshinski-Moriya interaction (DMI) term, the magnetic Skyrmion with a non-zero winding and finite size appear as the lowest energy topological configuration or ground state configuration~\cite{nnano.2013.243, Han:2017fyd}. In this work, we extend the conventional {nonlinear sigma model}  by coupling it to a gauged Ginzburg-Landau theory. This coupling happens via a spin-magnetic field interaction (SMFI). As we demonstrate through the Euler-Lagrange equations, this coupling leads to a novel hybrid structure in which the vortex configuration of the superconducting order parameter is intrinsically linked to a Skyrmion-like spin texture. In particular, within the region influenced by the vortex-induced magnetic field, the spin field exhibits a Skyrmion-like configuration. This interplay becomes evident from the dependence of the Skyrmion radius (as defined in Sec.~\ref{section-I}) on the characteristic length scale of the vortex, namely the penetration depth of the superconducting component. Beyond this region, the spin field exhibits oscillatory behavior as a function of radial distance. Furthermore, we find that the nature of the solution is sensitive to the sign of the vortex winding number. Specifically, a well-defined Skyrmion-like structure emerges only when the central spin aligns with the vorticity direction dictated by the sign of the vortex winding number $N$. We discuss the Skyrmion-vortex composite structure in absence and presence of DMI term. We also provide an analysis on the stability of such composite structure by minimizing the energy of the system with respect to Skyrmion length scale and vortex winding number. This interplay between two different topological objects suggests intriguing physical consequences that may be observable in experimental systems.\\

Another interesting aspect of these systems is the behaviour of the spin waves in the bulk. 
The existence of ferromagnetic order as well as the domain structure in a superconductor 
are hidden due to the Meissner effect. Due to this, the usual methods of studying magnetic properties of a system like Knight 
shift or muon spin relaxation {are not useful}. However, spin waves provide a direct probe into the nature of bulk magnetization, 
its direction and degree of uniformity. This provides a strong motivation for studying spin waves in such systems. The 
first study along this direction was done by Ng and Varma\cite{PhysRevB.58.11624} and later by Braude and 
Sonin~\cite{PhysRevLett.93.117001}. They have shown that, unlike other magnetic systems, in such ferromagnetic 
superconductors the circularly polarized spin waves have two modes having both positive and negative group velocity. 
In our work we shall show that the frequency $\omega$ satisfy a cubic equation in general. To get a physical picture we shall try to solve the cubic equation with various approximations. We shall show that in the low frequency regime the dispersion relation derived in~\cite{PhysRevLett.93.117001} appears. Also, we find that the spin wave dispersion curve has more than one local minima. The excitations around the non-trivial minimum are similar to roton modes of superfluids. Finally, we have found the spin wave equation in presence of a static vortex along the $z$ direction -- it suggests that in this case, for a given $\omega$ and $k$, the spin wave amplitude depends on the distance from the vortex core.\\

The plan of the paper is as follows. In Sec~\ref{section-I} we introduce the model under consideration and through the solutions of Euler-Lagrange equation we try to argue about the coupled Skyrmion-vortex structure. In Sec~\ref{section-II} we introduce the DMI term and discuss how the solution of the spin field changes in presence of this term. In Sec~\ref{section-III} We have discussed on stability of the composite structure. In Sec~\ref{section-IV} we derive the spin wave dispersion relation for different frequency regimes and discuss interesting features of the dispersion curve. Finally, in Sec~\ref{section-V} we conclude with some discussion on our results and directions for future research.
 
 
\section{nonlinear sigma model coupled to Ginzburg-Landau model}\label{section-I}
\begin{figure}[H]
    \centering
    \includegraphics[width=8cm,height=6cm]{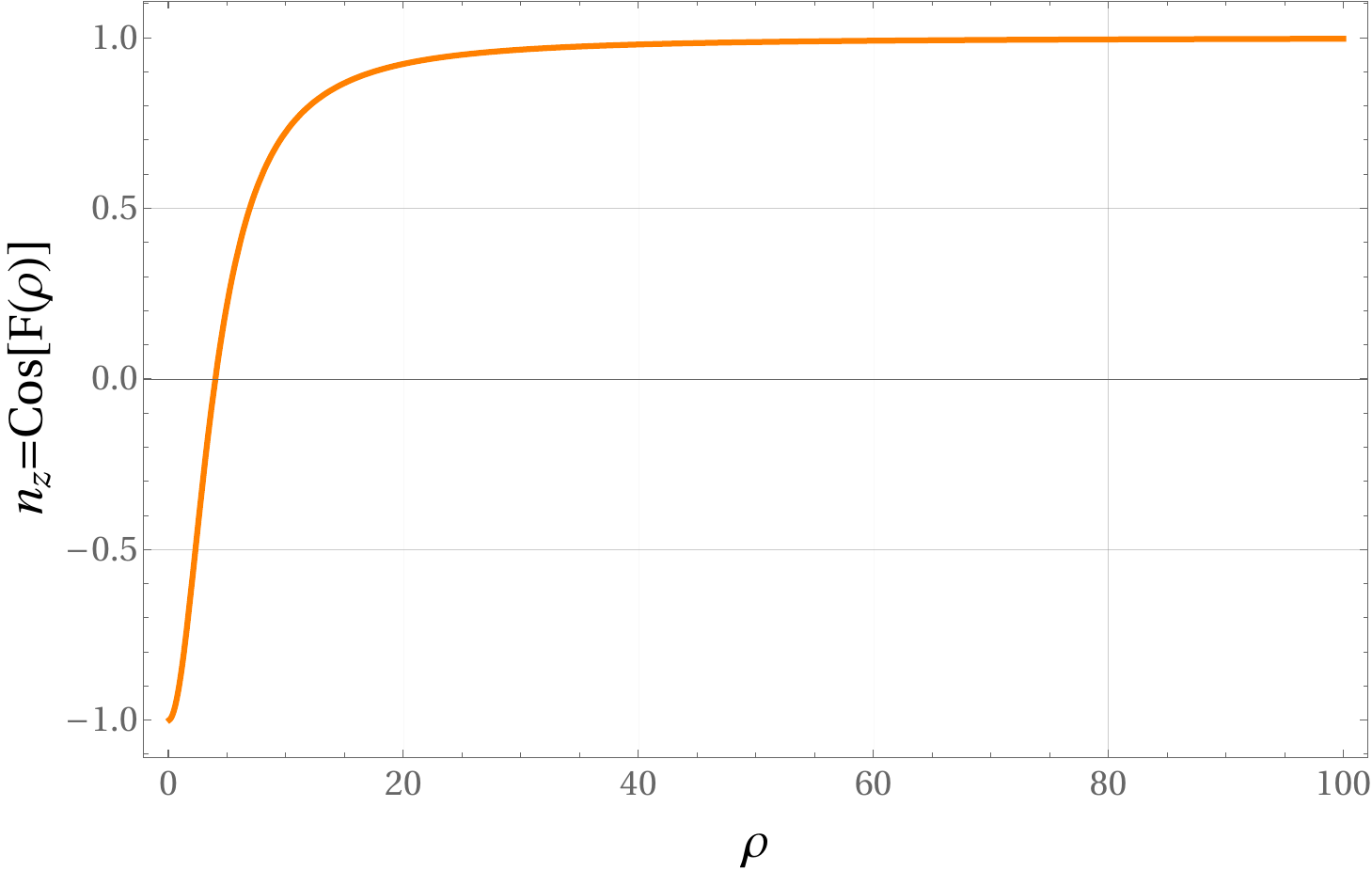}
    \caption{Skyrmion solution}
    \label{fig:1}
\end{figure}

We start by discussing Skyrmion solutions in the nonlinear sigma model~\cite{PhysRevD.52.2891} which can be expressed by the following Lagrangian, 
\begin{equation}
\mathcal{L}_{\sigma}= \frac{\rho_M}{2}\left(\mathbf{\grad} n^i\right)^2,
\end{equation}
where $\rho_M$ is called the ground state spin stiffness and is defined as $\rho_M= JS^2 a^{2-d}$. Here $J$ is the coefficient of the exchange coupling term of the microscopic lattice model, $a$ is the lattice spacing and $S$ is spin per site\cite{PhysRevLett.75.3509}.  $\mathbf{\hat{n}}$ is the unit vector along the magnetization density ${\mathbf M}$ whose magnitude $|\mathbf{M}|= \mu$ is assumed to be a constant in this model. This model can be obtained in the continuum limit of the Heisenberg model and it describes a magnetic system where the amplitude of the magnetic moments is fixed while the direction is varying in space. To derive the Skyrmion solution from this model we write the unit vector $\mathbf{\hat{n}}$ as
\begin{equation}\label{skyr. ansatz}
n^x= \sin F(\rho) \cos \phi,\, n^y= \sin F(\rho) \sin \phi,\, n^z = \cos F(\rho), 
\end{equation}
where $(\rho, \phi, z)$ are cylindrical coordinates in space, and $F(\rho)$ represents the angle of spin field $\mathbf{\hat{n}}$ with the $z$-axis at a radial distance $\rho$ from the origin. Eq.~\eqref{skyr. ansatz} defines a mapping between two dimensional unit sphere in spin space and two dimensional plane perpendicular to the Skyrmion axis. We proceed by rewriting the Lagrangian using the ansatz of Eq.~\eqref{skyr. ansatz} and the resulting Lagrangian is
\begin{equation}
\mathcal{L}= \frac{\rho_M}{2} \left(\partial_i F \partial_i F  + \frac{1}{\rho^2}\sin^2 F\right).
\end{equation}
Now we derive the Euler-Lagrange equation for the field $F(\rho)$ from the above Lagrangian and find the equation
\begin{equation}\label{skyr. sol.}
\partial_\rho^2 F + \frac{1}{\rho} \partial_\rho F - \frac{\sin 2F}{2\rho^2}=0.
\end{equation}
The solution of this equation corresponding to the initial condition that $F(\rho=0)=\pi,\, \partial_\rho F(\rho=0)=1$ behaves as shown in Fig.~\ref{fig:1}\,.

\begin{figure}[H]
    \centering
    \includegraphics[width=9cm, height=6cm]{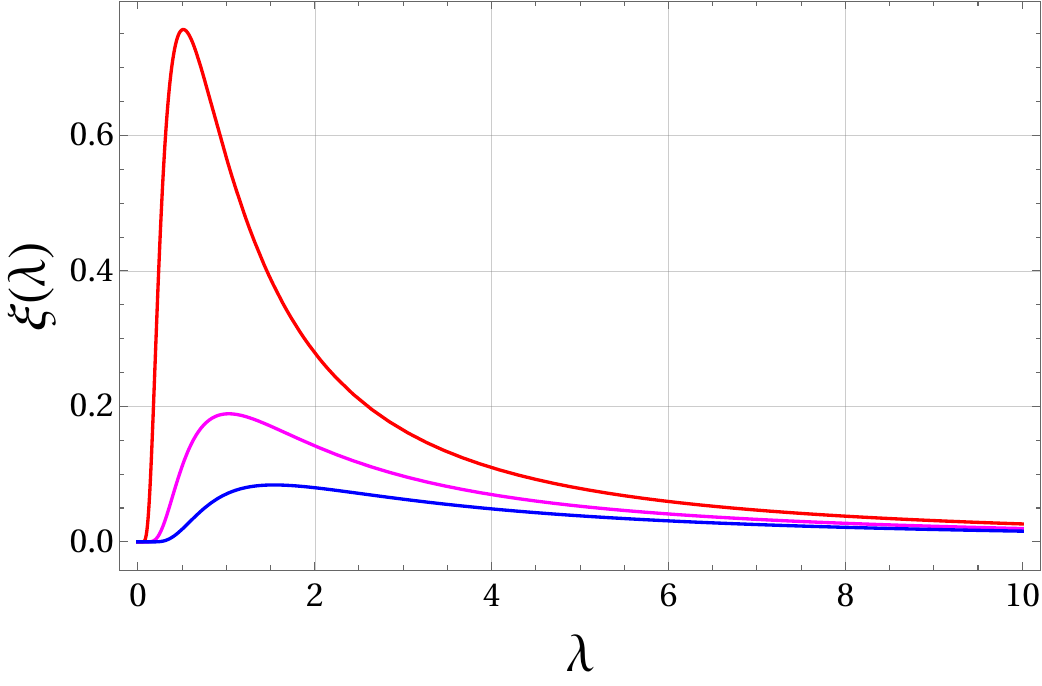}
    \caption{The function $\xi$ for three different values of distance $\rho$: $\rho$=2 (Red), $\rho$=6 (Magenta), $\rho$=10 (Blue)}
    \label{fig:2}
\end{figure}

\begin{figure}[H]
  \centering
  \subfloat[]{\includegraphics[width=0.45\linewidth]{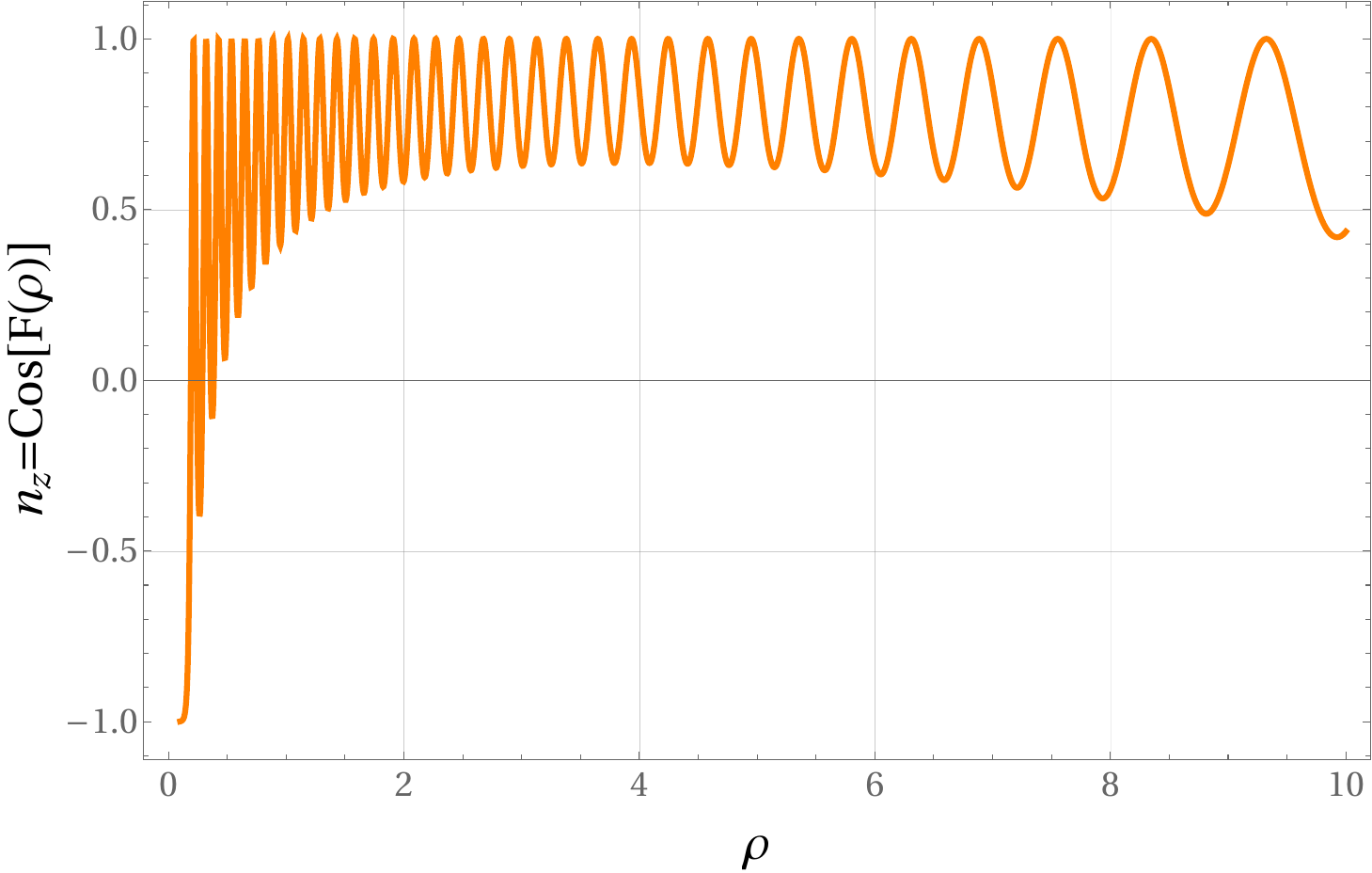}}\qquad 
    \subfloat[]{\includegraphics[width=0.45\linewidth]{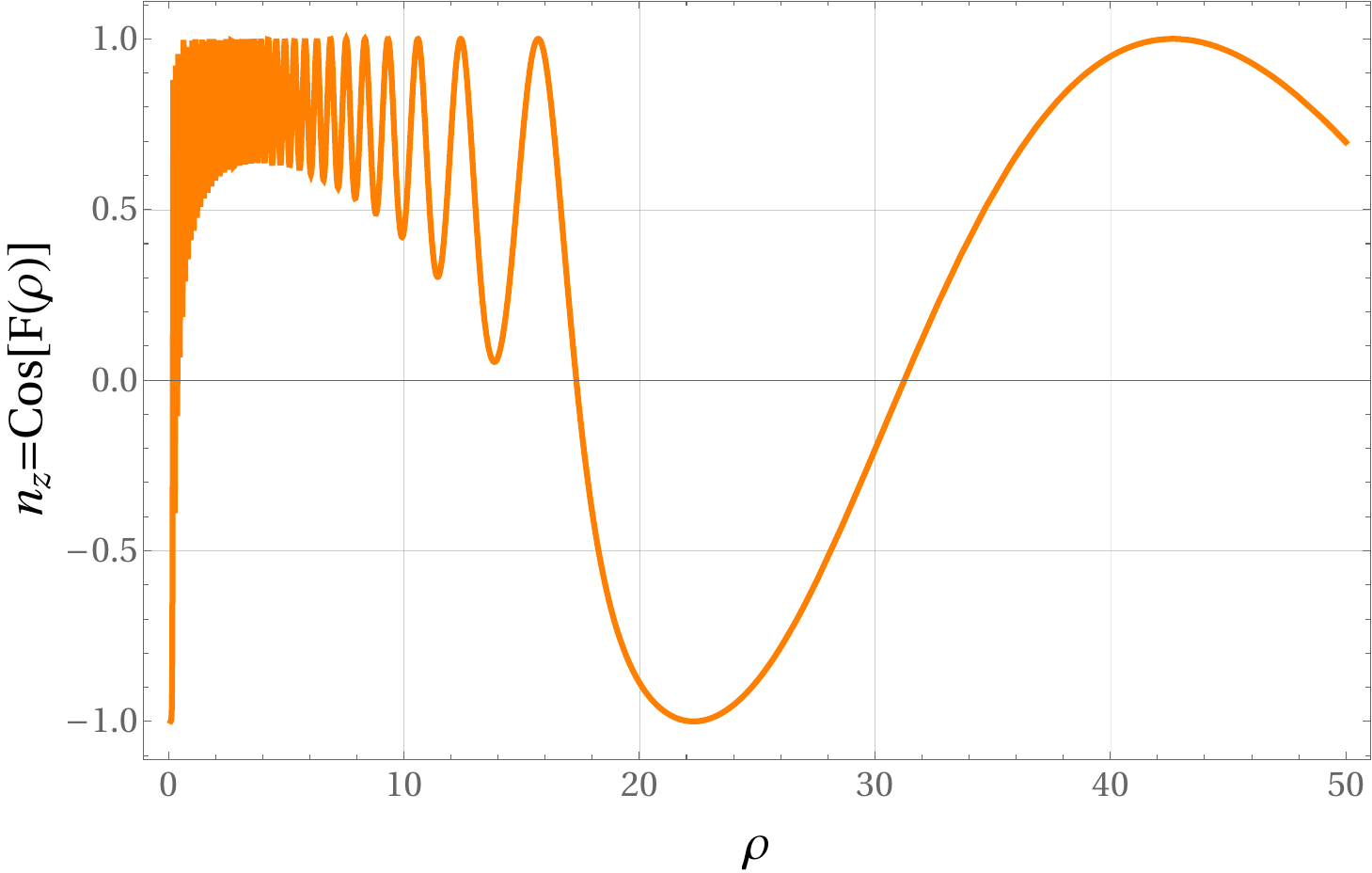}}\,
   \caption{Skyrmion solution for $\lambda=1$ eV$^{-1}$: a) short range behavior, b) long range behavior.}
  \label{fig:3}
\end{figure}

\begin{figure}[H]
    \centering
    \subfloat[]{\includegraphics[width=0.45\linewidth]{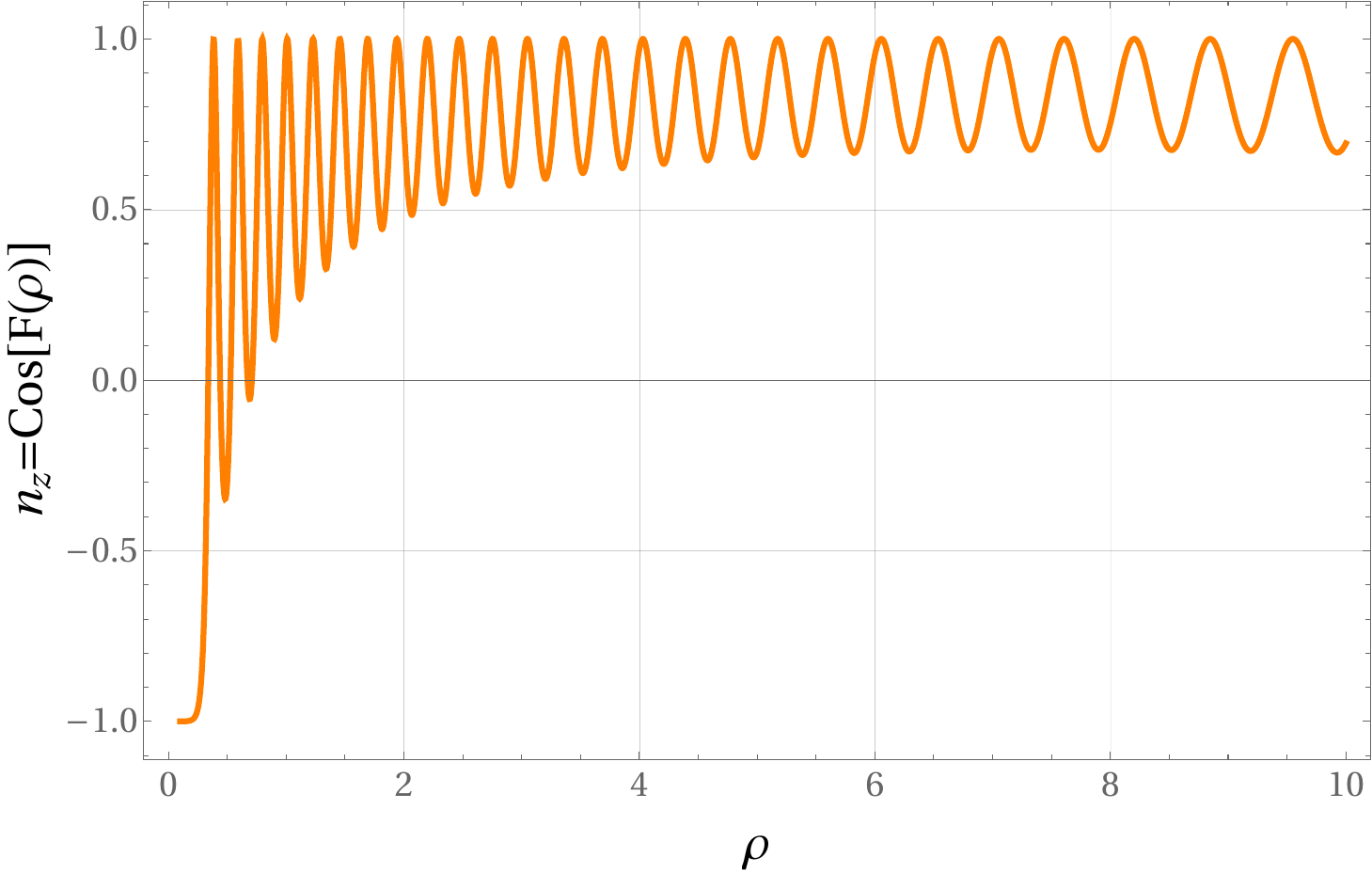}}
    \qquad
   \subfloat[]{\includegraphics[width=0.45\linewidth]{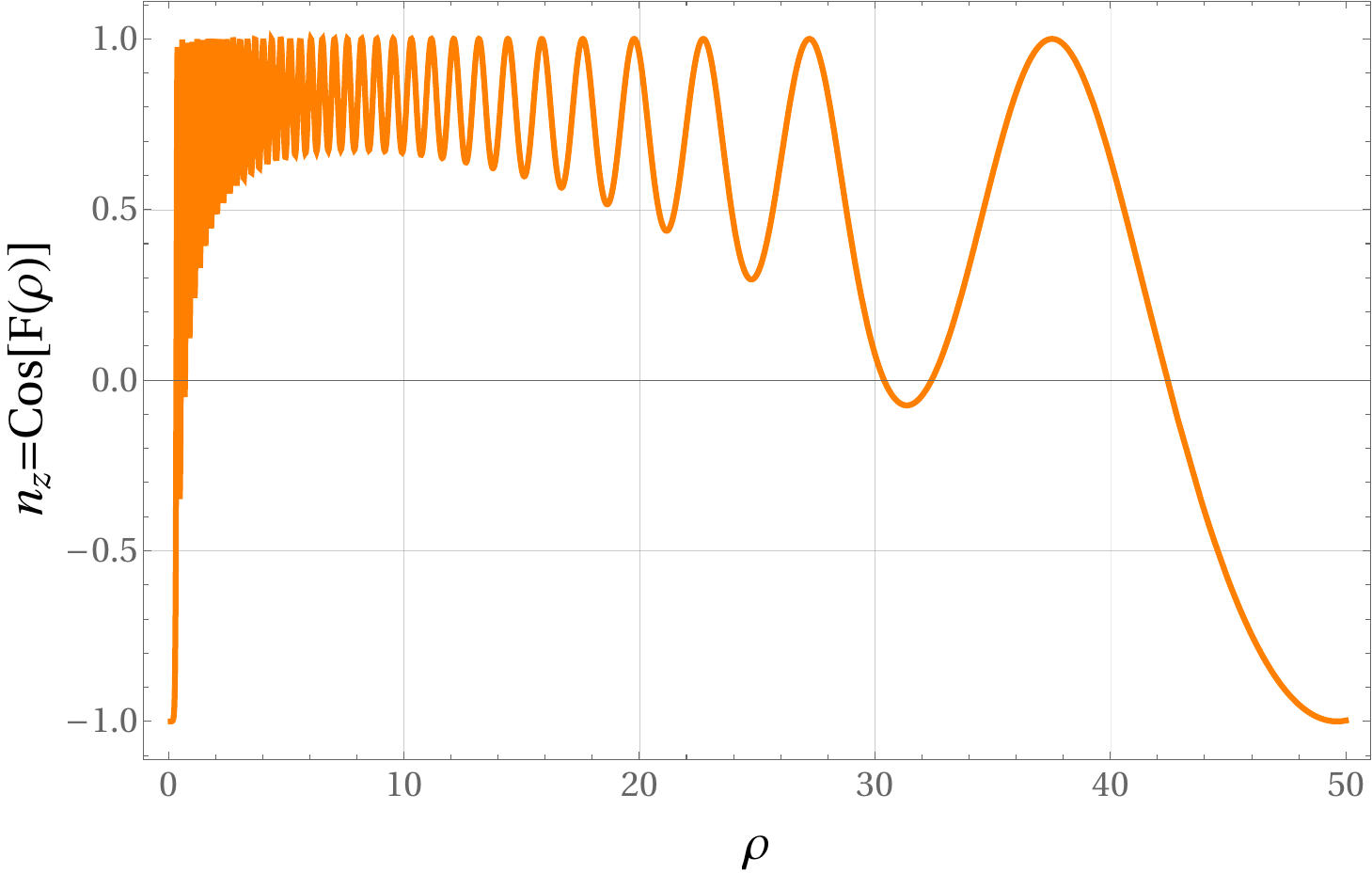}}\,
    \caption{Skyrmion solution for $\lambda=2$ eV$^{-1}$: a) short range behavior, b) long range behavior.}
    \label{fig:4}
\end{figure}

\begin{figure}
     \centering
     \includegraphics[width=10cm, height=6cm]{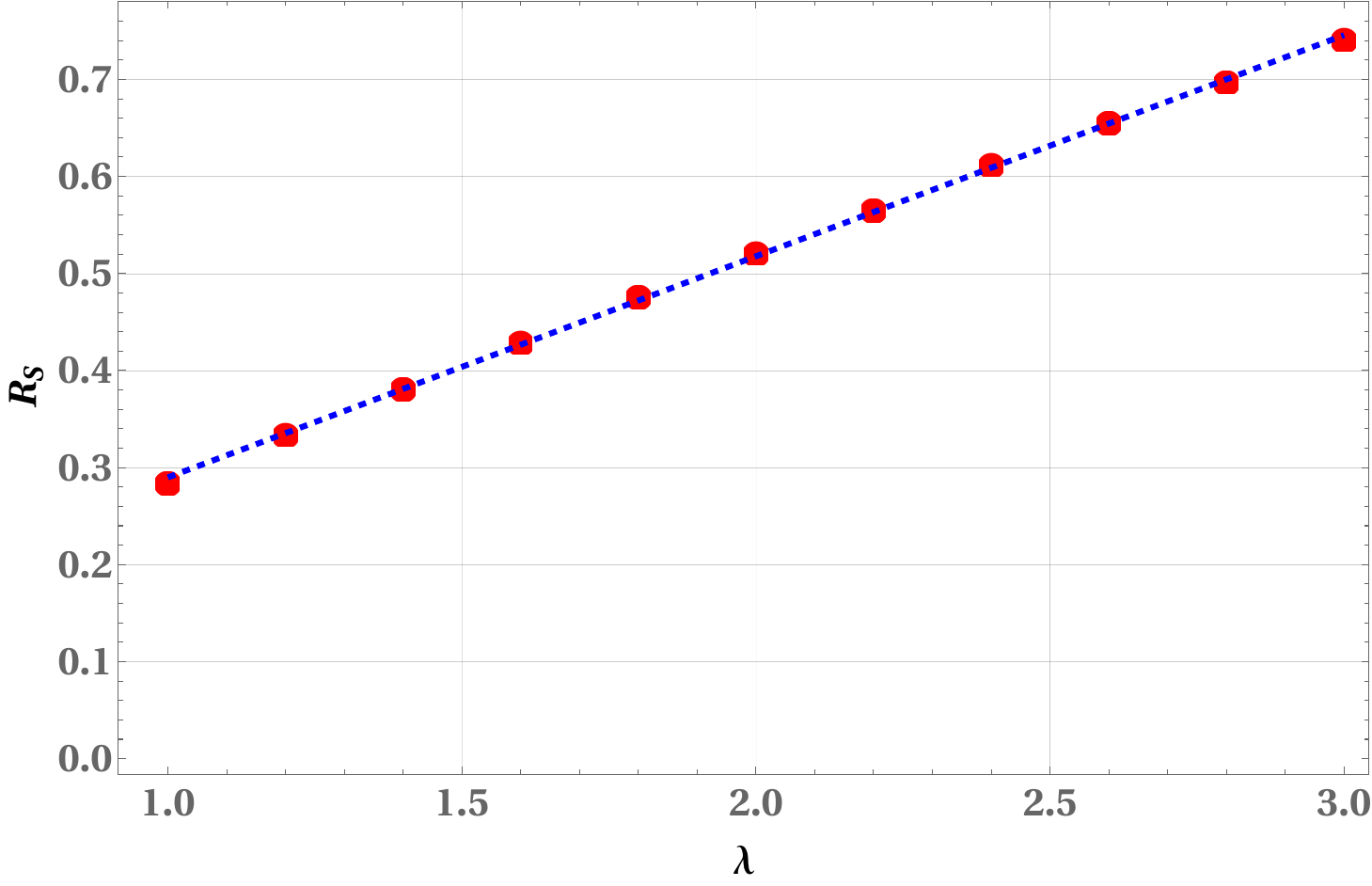}
     \caption{Skyrmion radius $R_{S}$ vs penetration depth $\lambda$ (both in eV$^{-1}$).}
     \label{fig:5}
 \end{figure}

It is known that different initial values of $\partial_\rho F$ give rise to Skyrmions of different sizes or radii~\cite{2020NatRP...2..492B}.
Skyrmions of this type are known as N\'eel type Skyrmions and their polarity is defined as~\cite{wang2018theory}
\begin{equation}
    p= \frac{n_z(r=\infty)- n_z(r=0)}{2}.
\end{equation}
Thus the polarity of this Skyrmion solution can be $p=\pm 1$. To describe a ferromagnetic superconductor, where the topological objects like Skyrmion and vortex interact, one needs to extend the former model suitably so that the ferromagnetic order would interact with the superconducting order. We shall consider an interaction term which couples the spin field $\mathbf{\hat{n}}$ with any magnetic field arising in the system. The resulting model is expressed by the following Lagrangian
\begin{equation}\label{nls-sup}
\mathcal{L}= -\frac{1}{4} F_{ij}^2 - \frac{\rho_M}{2}\left(\mathbf{\grad} n^i\right)^2 -\frac{\rho_s}{2m}(\mathbf{\grad}\theta - q\vec{A})^2 -\mu n^i \varepsilon^{ijk} \partial^j A^k.
\end{equation}
Here $\rho_s$ represents the superfluid density, $q$ and $m$ are the charge and mass of the Cooper pairs and $\mu$ is the amplitude of  magnetization, $\mu=|\mathbf{M}|$. Here we have considered the limit in which the amplitude of the superconducting order parameter is constant while the phase field $\theta$ varies in space. A winding in the phase variable introduces topological vortices into the theory. The core of the vortex contains a normal region where the order parameter vanishes. Here we shall consider this core radius to very small or effectively zero in comparison to penetration depth $\lambda$. The unit magnetization vector $\mathbf{\hat{n}}$ is assumed to vary over a unit sphere in spin space and therefore described by Eq.~\eqref{skyr. ansatz}. Putting this ansatz into the Lagrangian~\eqref{nls-sup} we derive the Euler-Lagrange equation for the fields $F(\rho)$ and $\vec{A}$. These are given by
\begin{align}
    \nabla^2 F - \frac{\sin 2F}{2\rho^2} - \frac{\mu}{\rho_M} \left(\sin F B^z - \cos F \cos\phi B^x- \cos F \sin\phi B^y \right)=0\,, \label{mod. Skyr.} \\
\left(-\nabla^2 + \frac{q^2\rho_s}{m}\right) \mathbf{B}= \frac{q\rho_s}{m} \pmb{\Sigma} -\mu \nabla^2 \mathbf{\hat{n}}\,.\label{EL B}
\end{align}
Here we have defined the vorticity $\Sigma^i= \varepsilon^{i j k}\partial^j \partial^k \theta_s$, where $\theta_s$ is the part of the phase which is multivalued and thus creates vortices in the system. The quantity $\sqrt{\dfrac{q^2\rho_s}{m}}$ acts as an effective mass of the photon. As suggested by Eq.~\eqref{EL B}, in the absence of superconductivity ($\rho_s=0$) the magnetic field is generated by spin configuration present in the system. In the presence of superconductivity this magnetic field may nucleate vortices in the system~\cite{PhysRevB.99.014511, PhysRevLett.46.49, PhysRevB.58.11624}. Here we consider that a vortex is created by the spin configuration and try to see what type of spin configuration is compatible to a vortex configuration. For that we first try to solve Eq.~\eqref{EL B} using the method of Green functions. The solution for $\mathbf{B}$ can be written as
\begin{equation}\label{B Sol}
    B^i(x) = \int d^3x^\prime \,\, G(\vec{x}-\vec{x}^\prime) \left[\frac{q\rho_s}{m} \Sigma^i -\mu \nabla^2 n^i\right] (\vec{x}^\prime)\,,
\end{equation}
where $G(\vec{x}-\vec{x}^\prime)$ is the Green function
satisfying the equation
\begin{equation}
    \left(-\nabla^2 + \frac{q^2\rho_s}{m}\right)G(\vec{x}-\vec{x}^\prime) = \delta^3(\vec{x}-\vec{x}^\prime)\,.
    \label{3d.Green}
\end{equation}
We can put this solution of Eq.\eqref{B Sol} into Eq.~\eqref{mod. Skyr.} to get a nonlinear integro-differential equation for $F$ (or $\pmb{\hat{n}}$) in terms of the sources $\pmb{\Sigma}$ and $\pmb{\hat{n}}$\,.
%
\begin{equation}\label{mod. Skyr.2}
\begin{aligned}
\nabla^2 F &- \frac{\sin 2F}{2\rho^2} - \frac{\mu}{\rho_M} \sin F \int d^3x^\prime \,\,  G(\vec{x}-\vec{x}^\prime) \left[\frac{q\rho_s}{m} \Sigma^z -\mu \nabla^2 n^z\right](\vec{x}^\prime) \\
& \qquad - \frac{\mu}{\rho_M}\cos F \cos\phi \int d^3x^\prime \,\,  G(\vec{x}-\vec{x}^\prime) \left[\frac{q\rho_s}{m} \Sigma^x -\mu \nabla^2 n^x\right](\vec{x}^\prime) \\ 
&\qquad - \frac{\mu}{\rho_M}\cos F \sin\phi \int d^3x^\prime \,\,  G(\vec{x}-\vec{x}^\prime) \left[\frac{q\rho_s}{m} \Sigma^y -\mu \nabla^2 n^y\right](\vec{x}^\prime) =0\,.
\end{aligned}
\end{equation}
In principle one can solve that equation self-consistently to know the resulting spin configuration as well as the vortex configuration which will be intertwined. However, the equation is a nonlinear integro-differential equation which is difficult to solve in general. To gain physical understanding out of this equation one may choose a simple vortex configuration and solve for the spin configuration compatible with this, or one can solve for the superfluid velocity $\textbf{v}_s=\mathbf{\grad}\theta_s$ for a given spin texture like Skyrmion. In this work we shall limit ourselves to the first case and shall leave the second possibility for a later work. Now, to proceed we shall take a straight vortex with the vortex line or string aligned along the $z$ axis. This configuration is expressed by the following equation
\begin{equation}
    \mathbf{\Sigma}= 2\pi N \int dz^\prime \mathbf{\hat{z}} \,\,\delta^3(\Vec{x}- z^\prime \mathbf{\hat{z}} ).
\end{equation}
This ansatz for the vortex configuration will simplify the Eq.~\eqref{mod. Skyr.2} by some degree. However, 
since $F(\rho)$ itself appears in the integrals of Eq.~\eqref{mod. Skyr.2}, 
{the} only possible way to proceed is to solve the equation in an iterative manner. For that we shall ignore {the $\nabla^2 n^i$ terms }
in Eq.~\eqref{mod. Skyr.2} and shall try to solve the remaining part. With all these in mind let us write down the approximated equation
\begin{equation}
    \nabla^2 F - \frac{\sin 2F}{2\rho^2} - \frac{\mu}{q \rho_M} \Tilde{m}^2 \sin F \int d^3x^\prime \,\,  G(\vec{x}-\vec{x}^\prime)  \Sigma^z(\vec{x}^\prime)=0,
\end{equation}
where we have written $\dfrac{q^2\rho_s}{m}= \Tilde{m}^2$ which is the photon mass in this case. Writing the superfluid density in natural units as $\rho_s= \dfrac{m}{2\pi q^2 \lambda^2}$, where $\lambda$ is the London penetration depth, we rewrite the above equation as
\begin{equation}
    \nabla^2 F - \frac{\sin 2F}{2\rho^2} - \frac{\mu}{2\pi q \rho_M\lambda^2} \sin F \int d^3x^\prime \,\,  G(\vec{x}-\vec{x}^\prime)  \Sigma^z(\vec{x}^\prime)=0,
\end{equation}
Recognizing the Green function $G(\vec{x}-\vec{x}^\prime)$ for  the differential operator $\left(-\nabla^2 + \dfrac{q^2\rho_s}{m}\right)$ in three spatial dimensions to be the Yukawa potential of a point source, we can perform the integration over the Green function as follows
\begin{equation}
\int dz^\prime \,\, G(\vec{x}-\mathbf{\hat{z}} \,\, z^\prime) =\int dz^\prime \frac{\exp\left(-\tilde{m}\sqrt{\rho^2 + (z-z^\prime)^2}\right)}{\sqrt{\rho^2 + (z-z^\prime)^2}}= \int_{-\infty}^\infty dx\,\, e^{-\tilde{m}\rho \cosh(x)}= 2K_0\left(\tilde{m}\rho\right).
\end{equation}
Finally we get the following simplified equation~\footnote[2]{An alternative way to derive this equation is to use a duality transformation, discussed in Appendix~\ref{Appendix II}.}
\begin{equation}\label{Effect of vortex on Skyr.}
     \partial_\rho^2 F +\frac{1}{\rho}\partial_\rho F - \frac{\sin 2F}{2\rho^2} - N \frac{2\mu}{q \rho_M\lambda^2} \sin F  K_0\left(\frac{\rho}{\sqrt{2\pi}\lambda}\right)=0.
 \end{equation}
We shall solve this equation numerically below~\footnote[1]{The numerical solution provided here have been evaluated for the parameter values specific to NbSe$_2$\cite{Zhu2016SignatureOC}. The details can be found in Appendix~\ref{Appendix I}}. It can be inferred by looking at {Eq.~(\ref{Effect of vortex on Skyr.})} above that the last term is the contribution from the vortex and may change the solution non trivially. However, for large $\lambda$ the last term will decouple from the rest and thus for extremely weak superconductor there will be no coupling between vortex and spin degrees of freedom. The same is also valid for very small penetration depth, {$\lambda\to 0\,,$} as the term starts to fall rapidly. This can be understood from the plot of the function $\xi =  \dfrac{1}{\lambda^2}  K_0\left(\frac{\rho}{\sqrt{2\pi}\lambda}\right)$ against $\lambda$ as we can see from Fig.~\ref{fig:2}.
It can be shown that in these extreme limits of penetration depth the last term has practically no effect on Skyrmion solution. {{However, if we use parameter values appropriate to, say NbSe$_2$, for which $\lambda\sim 1 $eV$^{-1}$ i.e. a few hundred nanometers, and $\dfrac{2\mu}{q \rho_M}\sim 63$~\footnotemark[1], a Skyrmion like solution exists within a region where the  magnetic field due to the vortex is non-vanishing.}} This is expressed by the plot of $n_z$ vs $\rho$ shown in Fig.~\ref{fig:3} and Fig.~\ref{fig:4}. 
%
{In addition, comparing Fig.~\ref{fig:3} and Fig.~\ref{fig:4} we can infer that with increment of penetration depth $\lambda$ the length of the region, over which Skyrmion like solution exists, also increases-- which indicates a coupling between these two objects.}
{To examine this coupling further,} one can also try to measure the changes in the Skyrmion radius due to change in $\lambda$. The Skyrmion radius $R_{S}$ is defined as a distance at which the sign of the $n_z$ changes. We have measured this distance for different values of $\lambda$. The variation of $R_{S}$ with penetration depth $\lambda$ is shown in Fig.~\ref{fig:5}\,.
This shows that Skyrmion radius is indeed affected by variation of penetration depth which in turn leads us to infer that in this solution vortex structure is coupled to Skyrmion-like spin configuration. Also, the solution shown above in Fig.~\ref{fig:3} is only possible for a specific choice of vorticity $N$. For $N>0$ ($N<0$) the Skyrmion like solution exist only if we take the initial condition $n_z=+1$ ($n_z=-1$) at the origin. The reason for choosing this initial condition is that in the vortex core the magnetic field strength is maximum and would lead to maximal alignment of the spin field at this point. This shows another evidence of coupling of the two topological structures. At this point we must mention that for any choice other than this results in a Skyrmion solution at a much greater length scale than penetration depth and it does not show any feature of coupling to the vortex. This shows that for the other choices of the initial condition the two structures decouple. \\

\begin{figure}[H]
    \centering
    \includegraphics[width=10
    cm, height=6cm]{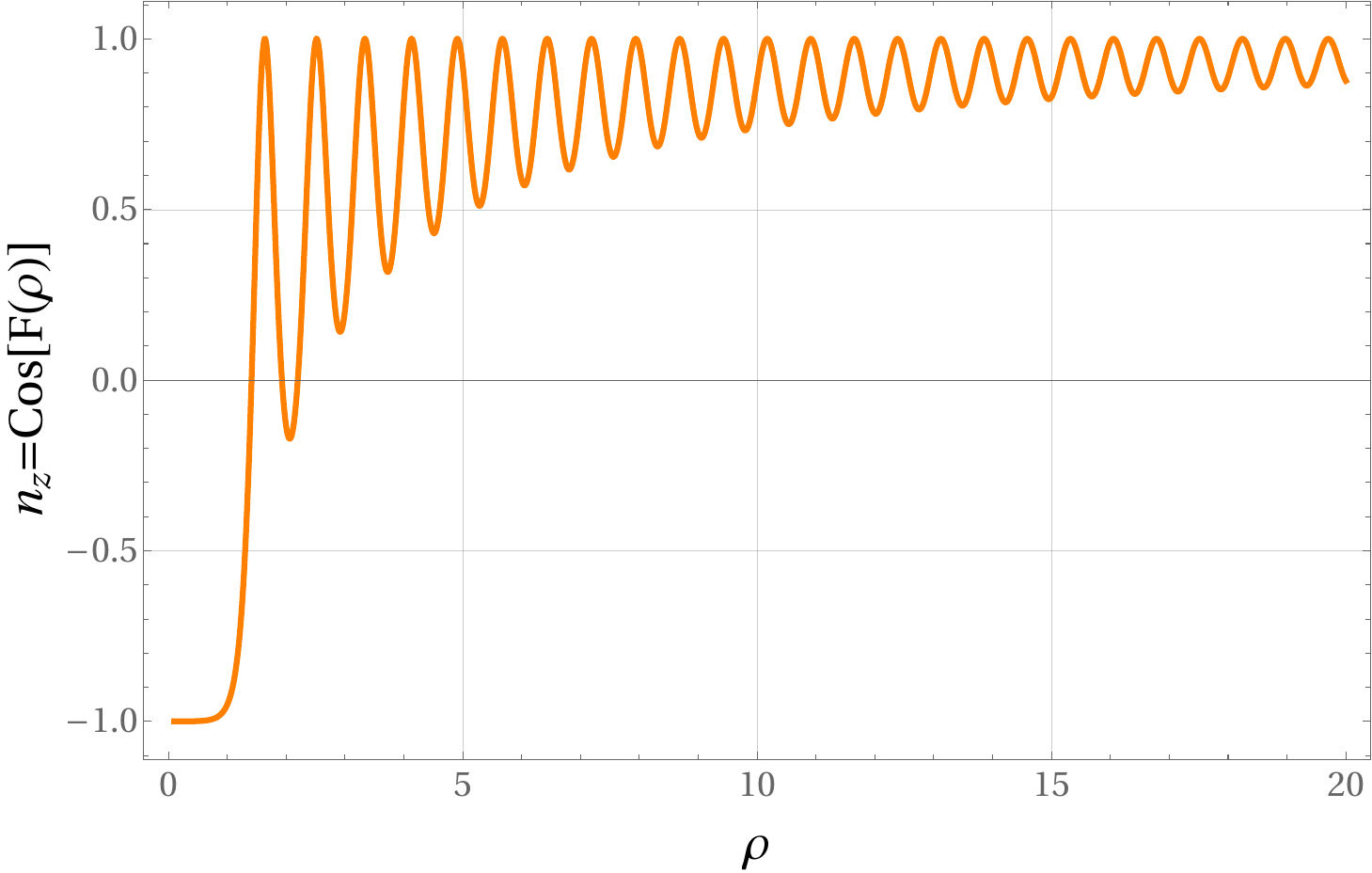}
    \caption{Skyrmion solution in a constant magnetic field}
    \label{fig:6}
\end{figure}

Thus we conclude that the Skyrmion-like configuration has appeared around the vortex configuration and extends only up to a finite region which is related to penetration depth of the superconductor. To understand the oscillating nature of the solution we note that in this region a finite magnetic field is present. Thus we first need to understand the nature of the Skyrmion in presence of a constant magnetic field. For that we take a look at the Eq.~\eqref{mod. Skyr.} and assume for a moment that the magnetic field is external and have components $\vec{B}=(0,0,B)$. For that we get the following equation for the field $F(\rho)$
\begin{equation}\label{mod. Skyr. const. mag. field}
\nabla^2 F - \frac{\sin 2F}{2\rho^2} - \frac{\mu B}{\rho_M}\sin F =0\,.
\end{equation}
We solve the above equation numerically for a magnetic field $B=-1 $ eV$^2$ and plot the solution in Fig.~\ref{fig:6}.
%
We see that the solution in this case looks exactly similar to that in the presence of a vortex. The only difference is that, in the case of a constant magnetic field, far from the centre the fluctuation becomes small and curve converges to $n_z=+1$ line. In comparison, in presence of a vortex, the fluctuation continues up to some length scale related to penetration depth and shows wave like oscillation outside that. This can be related to the fact that magnetic field in the system is penetrating only through the vortex and extends only in a finite region. On the other hand the wave like oscillating features outside this region might have some similarity to the oscillations in spiral magnetic phase discussed by Ng and Varma~\cite{PhysRevLett.46.49}.

\section{Dzyaloshinskii-Moriya Interaction and Bloch Skyrmion}\label{section-II}
\begin{figure}[H]
    \centering
    \subfloat[]{\includegraphics[width=0.45\textwidth]{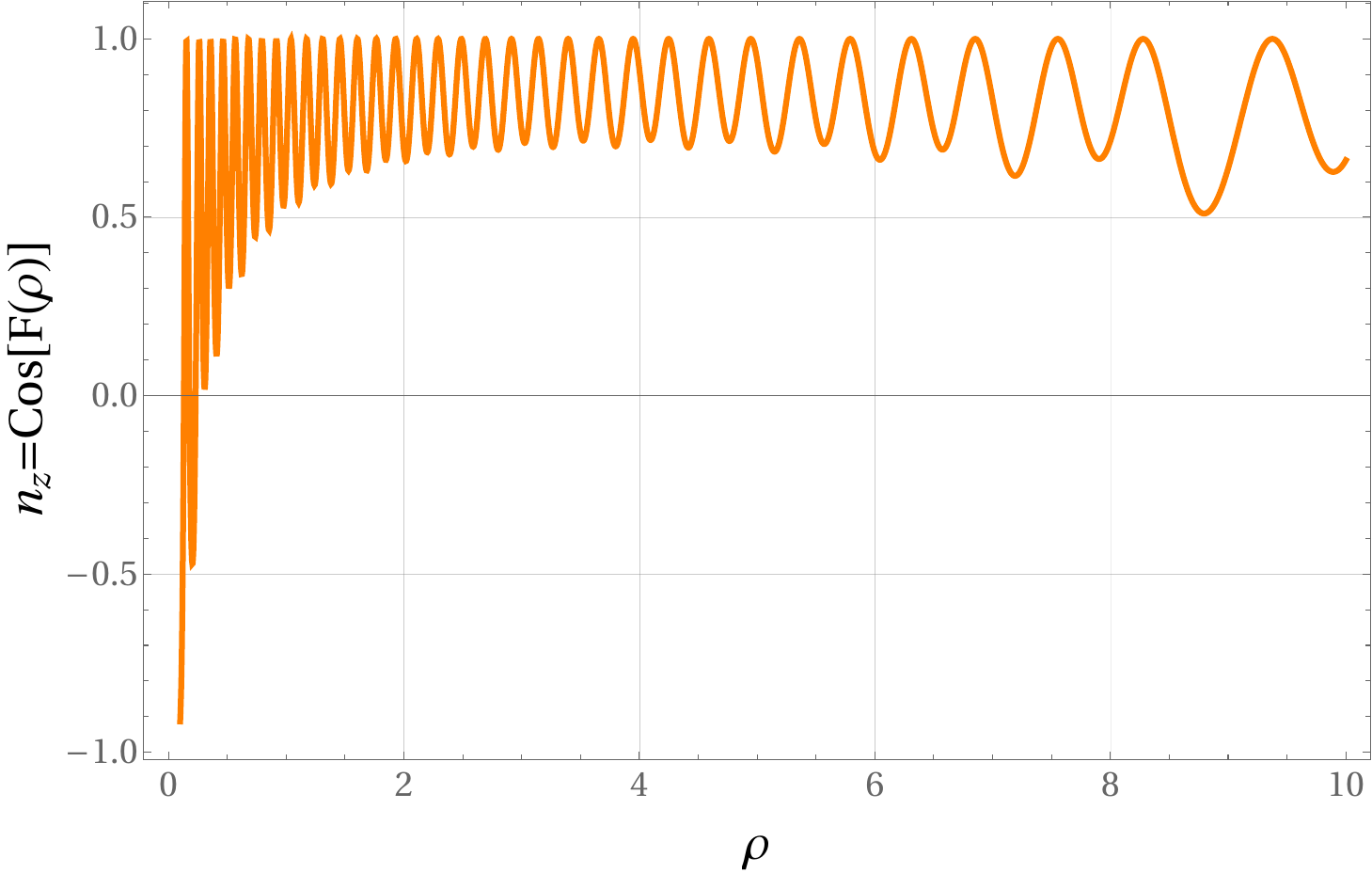}}
    \qquad
    \subfloat[]{\includegraphics[width=0.45\textwidth]{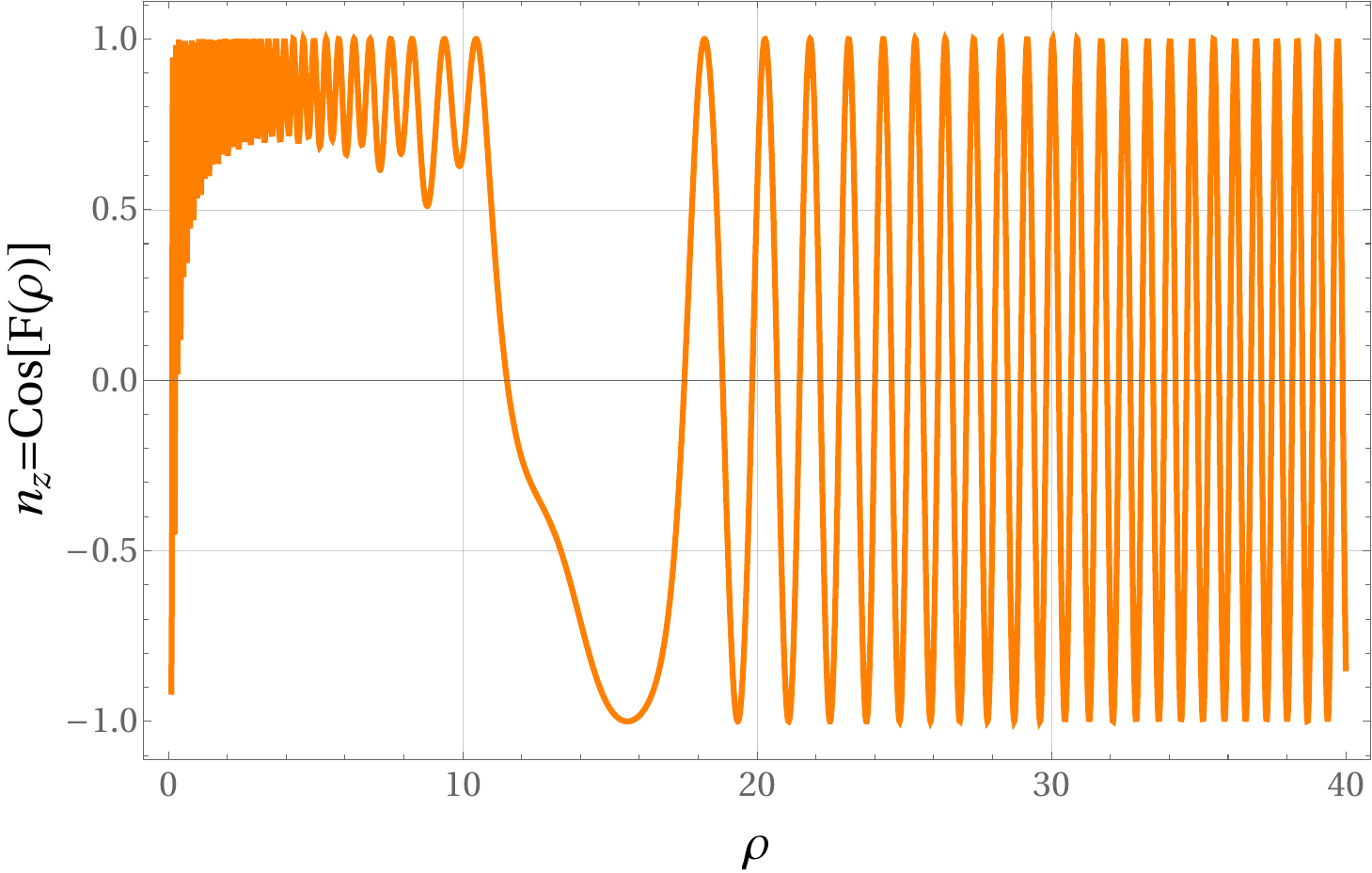}}
    \caption{Skyrmion solution with DMI interaction: a) short range behavior, b) long range behavior.}
    \label{fig:7}
\end{figure}

In the previous section we have shown how a coupled structure of vortex and Skyrmion-like spin configuration arises in {a nonlinear sigma model} coupled to {Ginzburg-Landau} theory of superconductivity. The Skyrmion configuration that we considered previously is the N\'eel type in which spins rotate in radial plane from core to the periphery. However, this configuration is one of the many possible configurations of Skyrmions which can be taken into account by a more general ansatz of the form
\begin{equation}\label{skyr. ansatz. gen.}
n^x= \sin F(\rho) \cos \Phi(\phi)\,, \; n^y= \sin F(\rho) \sin \Phi(\phi)\,,\; n^z = \cos F(\rho)\,.
\end{equation}
Here, the field $\Phi$ must be determined from the Euler-Lagrange equations. The general solution for the field $\Phi$ in the absence of the DMI term is of the form $\Phi= m\phi+\gamma$, where $m$ is the winding number for Skyrmion and the phase $\gamma$ decides the helicity. As we shall see, in presence of DMI the solution takes the specific form  $\Phi= \phi \pm \dfrac{\pi}{2}$. Furthermore, this particular configuration becomes the lowest energy (or ground state) configuration in presence of the DMI term. For this reason, the DMI term is said to stabilize the Skyrmion configuration\cite{nnano.2013.243}. We shall try to discuss these points in the context of a nonlinear sigma model in presence of a DMI term which is expressed by the Lagrangian
    \begin{equation}\label{nls-dmi1}
\mathcal{L}=  - \frac{\rho_M}{2}\left(\nabla n^i\right)^2 - D \varepsilon^{ijk} n ^i \partial^j n^k.
\end{equation}
With the above generalized ansatz, Eq.~\eqref{nls-dmi1} becomes
\begin{equation}\label{nls-dmi2}
\begin{aligned}
    \mathcal{L}= - \frac{\rho_M}{2}\left(\partial_i F \partial_i F  + \frac{1}{\rho^2} \left(\frac{\partial \Phi}{\partial \phi}\right)^2\sin^2 F\right) -D \left(\frac{\partial F}{\partial \rho} + \frac{1}{2\rho}\frac{\partial \Phi}{\partial \phi}\sin 2F\right) \sin (\Phi - \phi)\,.
    \end{aligned}
\end{equation}
The Euler-Lagrange equation for the fields $F$ and $\Phi(\phi)$ are then
\begin{equation}\label{mod. Skyr.DMI F}
\nabla^2 F - \frac{1}{2\rho^2}\left(\frac{\partial \Phi}{\partial \phi}\right)^2\sin 2F + \frac{D}{\rho_M \rho}\left(\frac{\partial \Phi}{\partial \phi}\right)\sin\left(\Phi-\phi\right)(\cos 2F -1) =0\,,
\end{equation}
and 
\begin{equation}\label{mod. Skyr.DMI Phi}
    \frac{1}{\rho^2} \frac{\partial^2 \Phi}{\partial \phi^2} - \frac{D}{2\rho \rho_M} \sin 2F \cos(\Phi -\phi) -\frac{D}{\rho_M} \frac{\partial F}{\partial \rho} \cos(\Phi-\phi)=0\,.
\end{equation}
One can check that the Eq.~\eqref{mod. Skyr.DMI Phi} is trivially satisfied if we take $\Phi= \phi \pm \dfrac{\pi}{2}$. This choice also leads to minimum energy Skyrmion configuration as shown by Nagaosa and Tokura \cite{nnano.2013.243} and this configuration is called a Bloch type Skyrmion. Putting this solution for $\Phi$ into Eq.~\eqref{mod. Skyr.DMI F} we get
\begin{equation}\label{mod. Skyr.DMI F 2}
\nabla^2 F - \frac{1}{2\rho^2}\sin 2F + \frac{D}{\rho_M \rho}(\cos 2F -1) =0.
\end{equation}
This equation leads to the desired Skyrmion solution in presence of a potential term as shown in~\cite{2020NatRP...2..492B}~(see Appendix~\ref{Appendix III}). With this background, we extend the model of Eq.~(\ref{nls-dmi1}) by coupling it to Ginzburg-Landau theory. The new model is expressed by the Lagrangian
\begin{equation}\label{nls-sup-dmi1}
\mathcal{L}= -\frac{1}{4} F_{ij}^2 - \frac{\rho_M}{2}\left(\mathbf{\grad} n^i\right)^2 -\frac{\rho_s}{2m}(\vec{\nabla}\theta - q\vec{A})^2 -\mu n^i \varepsilon^{ijk} \partial^j A^k - D \varepsilon^{ijk} n ^i \partial^j n^k\,.
\end{equation}
{Using the ansatz of Eq.~(\ref{skyr. ansatz. gen.}), we can write the } Euler-Lagrange equations for $F$ and $\Phi$ as
\begin{equation}\label{mod. Skyr.DMI F mag}
\nabla^2 F - \frac{1}{2\rho^2}\left(\frac{\partial \Phi}{\partial \phi}\right)^2\sin 2F + \frac{D}{\rho_M \rho}\left(\frac{\partial \Phi}{\partial \phi}\right)\sin\left(\Phi-\phi\right)(\cos 2F -1) - \frac{\mu}{\rho_M} \left(\sin F B^z - \cos F (\cos \Phi B^x+ \sin \Phi B^y)\right) =0.
\end{equation}
\begin{equation}\label{mod. Skyr.DMI Phi mag}
    \frac{1}{\rho^2} \frac{\partial^2 \Phi}{\partial \phi^2} - \frac{D}{2\rho \rho_M} \sin 2F \cos(\Phi -\phi) -\frac{D}{\rho_M} \frac{\partial F}{\partial \rho} \cos(\Phi-\phi)- \mu \sin F(\rho) (\sin\Phi(\phi) B^x +  \cos\Phi(\phi) B^y )=0\,.
\end{equation}
{We notice that the Euler-Lagrange equation for $\Phi$ in the present case becomes identical with Eq.~\eqref{mod. Skyr.DMI Phi} when the vortex is along the $z$ axis, since in that case $B^x=B^y=0\,, B^z\neq 0$. Thus the solution to the field equation for the field $\Phi$ is $\Phi=\phi \pm \dfrac{\pi}{2}$}. Putting the solution $\Phi=\phi + \dfrac{\pi}{2}$ to the Euler-Lagrange equation for $F(\rho)$ we get the following equation for $F$
\begin{equation}
    \nabla^2 F - \frac{1}{2\rho^2}\sin 2F  + \frac{D}{\rho_M \rho} (\cos 2F -1) - \frac{\mu}{\rho_M} \sin F\, B^z  =0
\end{equation}
We shall now proceed similarly to what was discussed in  Sec~\ref{section-I}. For a $z$-directional vortex configuration and in the first order approximation for the spin-spin interaction term, we have the following decoupled equation
\begin{equation}
    \nabla^2 F - \frac{1}{2\rho^2}\sin 2F + \frac{D}{\rho_M \rho}(\cos 2F -1)-  N \frac{2\mu}{\lambda^2 q \rho_M} \sin F  K_0(\frac{\rho}{\sqrt{2\pi}\lambda}) =0.
\end{equation}
The last term in the above equation due to the vortex provides the potential term needed for Skyrmion solution. However, this term decays with distance very rapidly and therefore the Skyrmion-like spin configuration is limited only up to a length scale which is determined by the penetration depth $\lambda$ and the ratio $\dfrac{2\mu}{ q \rho_M}$\,. Outside this length scale the solution exhibits spiral wave like configuration which is a feature of the solution in presence of the DMI term and in the absence of the potential terms. The numerical solution for the choices~\footnotemark[1] $\lambda= 1$ eV$^{-1}$, $\dfrac{D}{\rho_M}=15.12$~eV, $ \dfrac{2\mu}{ q \rho_M}=63$ and $N=-10$ is shown in Fig.~\ref{fig:7}. Thus the solution in presence of DMI shows very similar qualitative features to previous solutions without DMI {close to the vortex}.

\section{Stability of the composite object}\label{section-III}
In the previous sections we have discussed Skyrmion-vortex composite solutions and their interdependent features in the context of Ginzburg-Landau + nonlinear sigma model in absence and in presence of DMI term. However, for realizing such topological objects in real systems these objects should be stable in the sense that the energy of such configuration should minimize the total energy of the system. In the discussion below we provide the necessary analysis in which we shall minimize the energy of the system with respect to two relevant parameters --- $(a)$ length scale of the spin configuration, $(b)$~winding number of the vortex. 
\subsection{Derrick's theorem and length scale $R$ of Skyrmion like configuration}
It is well known in the context of physics of solitons that stability of a soliton solution is examined by Derrick's theorem~\cite{Manton:2004tk, Han:2017fyd}. It states that finite energy stable solutions of a field theory should correspond to the minima of energy expressed as a function of spatial scaling factor. {For the Skyrmion solution in non-linear sigma model the relevant length scale is the radius $R$ of the Skyrmion, defined as the distance from the center at which the $z$-component of the spin vanishes.} {To apply this theorem to our case, we need the energy functional for the system of interest which can be obtained from Eq~\eqref{nls-dmi2} using a Legendre transform
\begin{equation}
  \mathcal{H} =  \Pi_{\mathbf{\hat{n}}} \partial_t\mathbf{\hat{n}} - \mathcal{L}\,, 
\end{equation}
where $\mathcal{H}$ is the Hamiltonian corresponding to the Lagrangian $\mathcal{L}$ given in in Eq~\eqref{nls-dmi2}, and $\Pi_{\mathbf{\hat{n}}}$ is the conjugate momentum for the field $\mathbf{\hat{n}}$. For the static case when the first term becomes zero, and in presence of the SMFI interaction, we get the following energy functional}
%
\begin{equation}\label{energy-functional}
{ E=  \int_0^\infty u\, du\left[\pi \rho_M\left[(\partial_u F)^2 + \frac{1}{u^2}(\sin F)^2\right] + 2\, R\,\pi D  \left[ \partial_u F + \frac{\sin(2F)}{2u}\right] -2\pi \mu B\, R^2 \cos F\right].}
\end{equation}
Here we have written the earlier $F$, which  represents the angle of the spin with the $z$-axis, as a function of the dimensionless radial coordinate $u=\dfrac{\rho}{R}$\,. As the spin configuration is assumed to be independent of the $z$ {coordinate}, this expression in Eq.~\eqref{energy-functional} is the total energy per unit length along the $z$ direction.\\

For constant $B$, varying the energy with respect to $R$ one can show that the energy functional has a minimum at a finite length scale $R$ which is fixed by the constant magnetic field strength $B$ and DMI coefficient $D$. Thus the Skyrmion solution is stable in presence of DMI and SMFI~\cite{Han:2017fyd}. We shall now follow the same method to check whether the Skyrmion-vortex composite solutions are stable.
{For Skyrmion-vortex composite solutions, we write $B= B_{\text{vortex}}+ B_0$ in the energy functional of Eq~\eqref{energy-functional}\,, where }
$B_{\text{vortex}}$ is the magnetic field due to the vortex, with a spatial profile given by
\begin{eqnarray}
    B_{\text{vortex}} (u)=  \frac{2 N}{q \lambda^2}  K_0\left(\frac{u R}{\sqrt{2\pi}\lambda}\right)\,,
\end{eqnarray}
and $B_0$ is a constant background magnetic field. The purpose of putting this background field will become clear shortly. We also note that due to the rescaling $\rho= u R\,,$ a factor of $R$ has appeared in the argument of the modified Bessel function. Thus the contribution due the vortex in the last term of Eq.~(\ref{energy-functional}), when integrated over the dimensionless variable $u$\,, will have a very nontrivial dependence on $R$. The total energy is now given by 
\begin{eqnarray}
    E=   \pi \rho_M I_{\text{NL}} +  2\pi D  I_{\text{DM}} R - 2\pi R^2\left( \frac{2N\mu}{q\lambda^2} I_{\text{Vortex-Spin}} + \mu B_0 I_{B_0\text{-Spin}}\right) ,
\end{eqnarray}
{where $I_{\text{NL}}$, $I_{\text{DM}}$, $I_{\text{Vortex-Spin}}$, $I_{B_0\text{-Spin}}$ represent the nonlinear sigma model term, the DMI term, the energy of interaction between spin field $\mathbf{\hat{n}}$ and the magnetic field of the vortex, and the energy of interaction between the spin field $\mathbf{\hat{n}}$ and the background magnetic field $B_0$\,,
\begin{align}
    I_{\text{NL}} &= \int_0^\infty u\, du\,\left[(\partial_u F)^2 + \frac{1}{u^2}(\sin F)^2\right]\,, \\ I_{\text{DM}} &= \int_0^\infty u\, du\,\left[ \partial_u F + \frac{\sin(2F)}{2u}\right]\,,\\ I_{\text{Vortex-Spin}} &=\int_0^\infty u\, du\, \cos F\,K_0\left(\frac{u R}{\sqrt{2\pi}\lambda}\right)\,,\\ I_{B_0\text{-Spin}} &= \int_0^\infty u\, du\, \cos F\,.
\end{align}
}

\begin{figure}[H]
    \centering\subfloat[]{\includegraphics[width=0.3\linewidth]{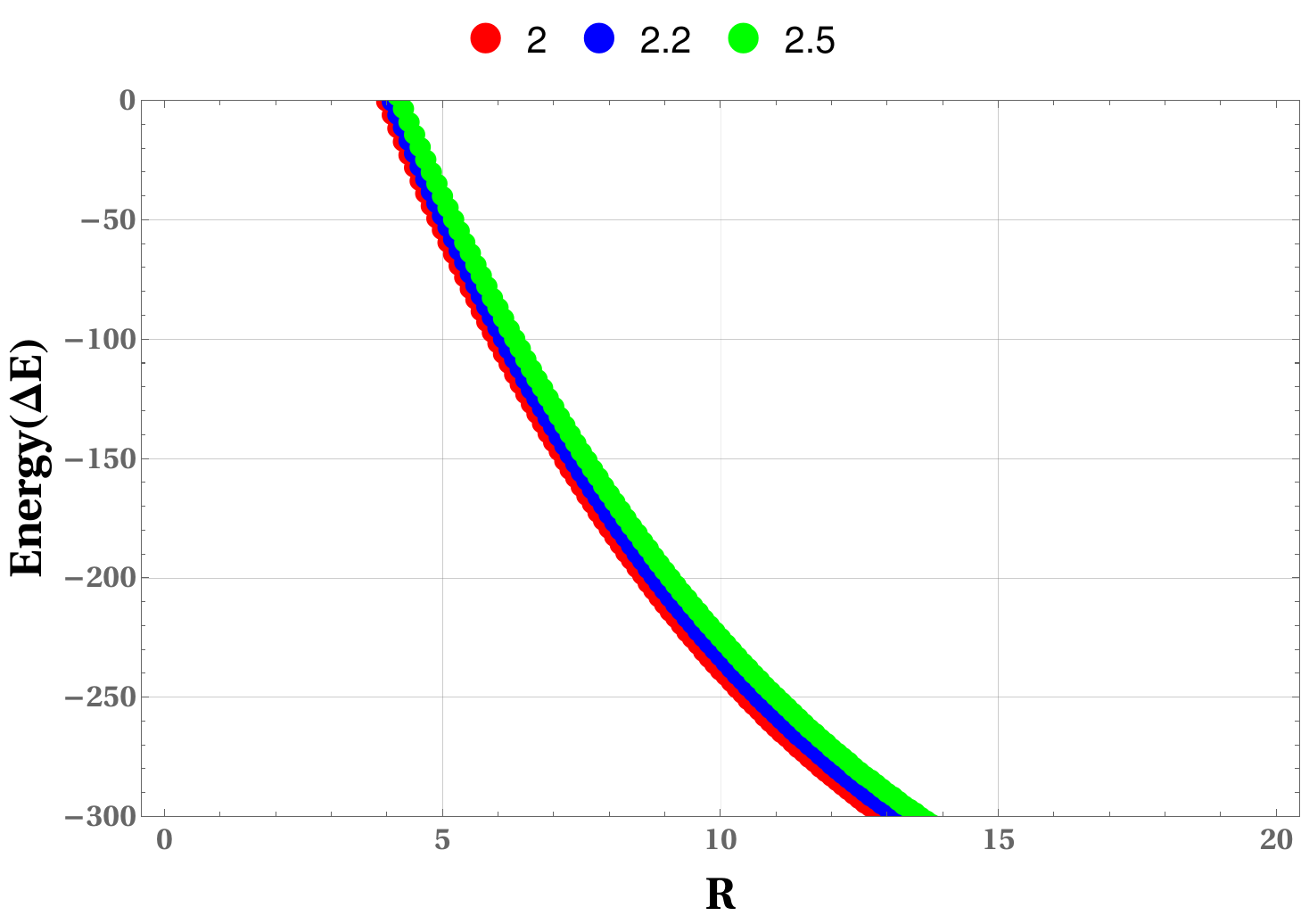}}\;
    \centering\subfloat[]{\includegraphics[width=0.3\linewidth]{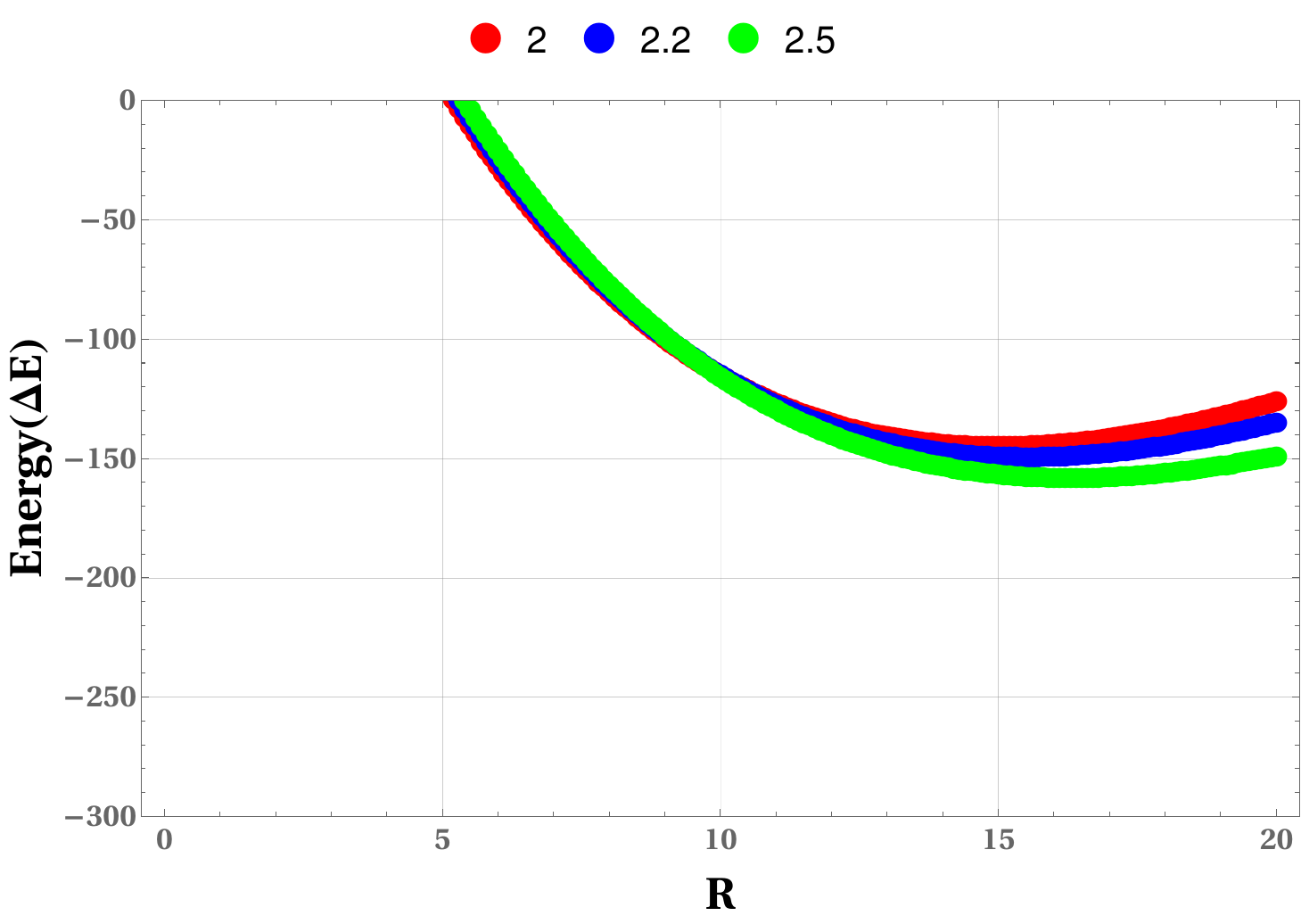}}\;
    \centering\subfloat[]{\includegraphics[width=0.3\linewidth]{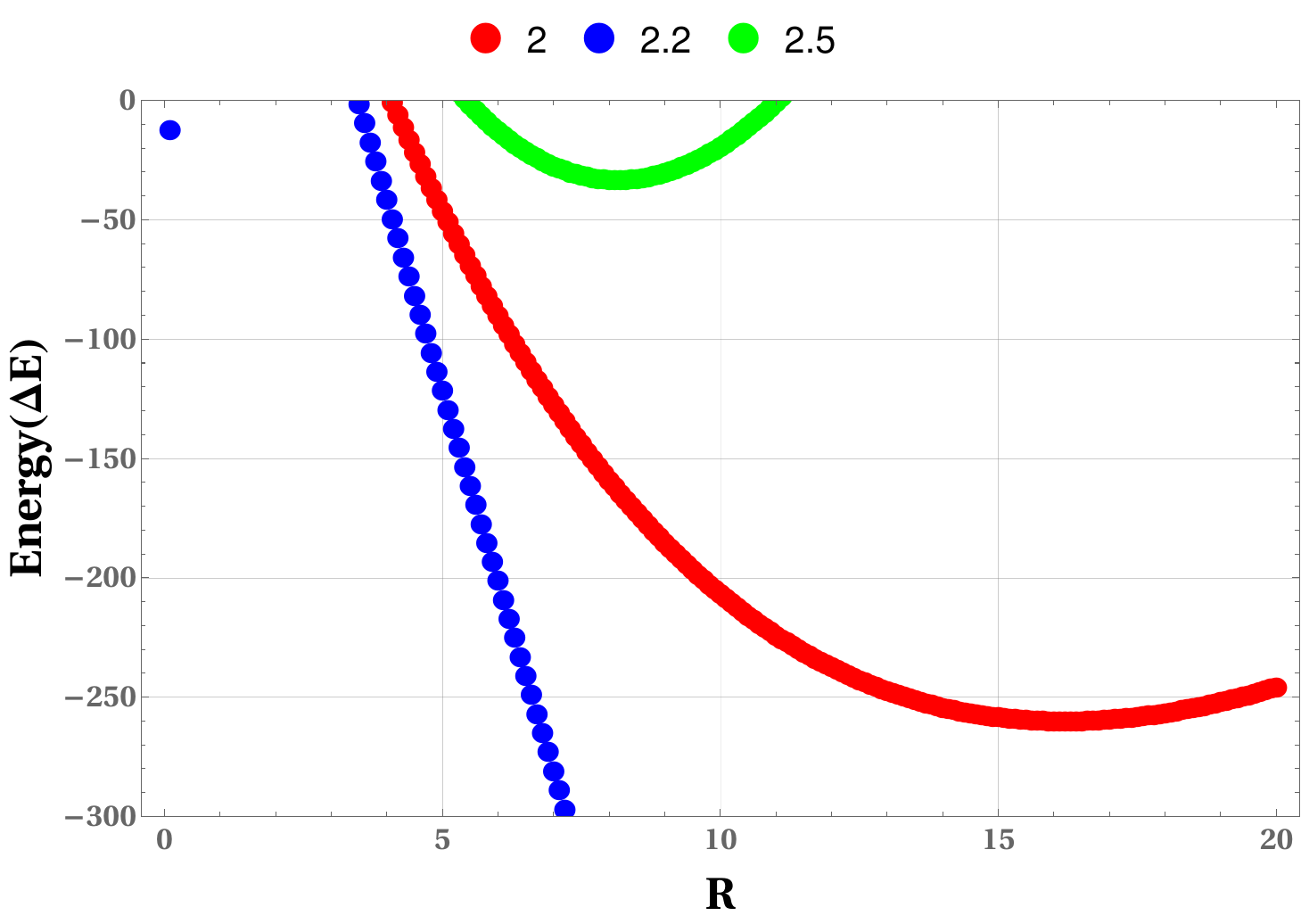}}
    \caption{Variation of $\Delta E$ (in eV) as a function of R (in eV$^{-1}$) for (a) $\tilde{D}=0,\, B_0=0$, (b) $\tilde{D}= -0.02\,\, \text{eV}^3,\, B_0=0$  ,(c) $\tilde{D}=0,\, B_0=0.001\,\text{eV}^2$ for $\lambda = 2, 2.2, 2.5$ (eV$^{-1}$).}
    \label{fig:8}
\end{figure}

Let us rewrite the total energy by redefining different parameters as 
\begin{align}\label{Energy_functional_final}
  \tilde{\mu} &=\frac{4\pi N\mu}{q\lambda^2}\,,\qquad \tilde{D}= 2\pi D\,,\qquad \mu^\prime= 2\pi \mu B_0\,,\\
\text{ so that } \qquad   E &= \pi \rho_M I_{\text{NL}}+ \tilde{D} I_{\text{DM}} R -  R^2\left( \tilde{\mu} I_{\text{Vortex-Spin}} + \mu^\prime I_{B_0\text{-Spin}}\right)\,.
\end{align}
We now evaluate the energy $\Delta E= E-  \pi \rho_M I_{\text{NL}}$ numerically by putting the numerical solution for $F(u)$ for different possible values of $\lambda$. Let us consider following three scenarios: $(a)$ the system lacks both the Dzyaloshinskii-Moriya interaction (DMI) and a background magnetic field, $(b)$ the system includes a DMI term with $D<0$ but no external magnetic field, and $(c)$ the system is subject to a finite external magnetic field but does not include a DMI term. The variation of $\Delta E$ vs. $R$ for these three scenarios are shown in Fig.~\ref{fig:8}.
In the absence of background magnetic field and DMI term if we solve the field equation for $F(u)$ and evaluate the energy functional for that solution numerically, we obtain the $\Delta E$ vs. $R$ plot shown in Fig.~\ref{fig:8}(a). This plot suggests that in the absence of the background field $B_0$ and the DMI term, no minimum appears and thus in this limit it is difficult to stabilize the solution. It is widely known that in the presence of the DMI term and an SMFI type term, a finite size Skyrmion solution stabilizes. We can also verify this for our model by considering the DMI term with a small negative DMI coefficient in the absence of any background magnetic field. This case is shown in Fig.~\ref{fig:8}(b) which shows that in this case a minimum indeed appears in the  $\Delta E$ vs. $R$ plot assuring the stability of the solution. In the third scenario we introduce a small background magnetic field $B_0=0.001$ eV$^2\sim 10^{-5}$T along the same direction as the vortex in the absence of the DMI term and obtain the  $\Delta E$ vs. $R$ variation shown in Fig.~\ref{fig:8}(c). We notice that in this scenario also a minimum has appeared. Thus we can conclude that either a DMI term or a background magnetic field $B_0$ is necessary to stabilize the Skyrmion solution. Although there is no evidence of a DMI-type interaction in a ferromagnetic superconductor phase, one can try to induce such a term by external means like doping of heavy elements~\cite{PhysRevMaterials.6.084401}, strain engineering~\cite{PhysRevLett.127.117204}, external electric field~\cite{2014NatSR...4.4105B} etc. On the other hand, an internally generated background magnetic field contribution along the direction of vortex is available in a vortex lattice environment which can also appear in a ferromagnetic superconductor phase. We also note that although the minima of energy appear at very large negative energies, adjustment of the nonlinear sigma term through manipulation of the exchange coupling constant $J$, structure constant $a$ etc. can bring the minima to physically realizable energies $\sim${meV}.

\subsection{Minimum value of winding number N}
Next we want to minimize the energy of the system with respect to $N$ to obtain the allowed values of $N$ for the composite excitation. To proceed we write down the terms contributing in total energy that explicitly depends on the winding number of the vortex in general
\begin{equation}\label{Total Energy N}
    E(N)= N E_{\text{int}} + N^2 E_{V} + N^2 E_{\text{VV}},
\end{equation}
where $E_{\text{int}}$ represents the interaction energy of the magnetic field configuration of a vortex and the Skyrmion-like spin configuration, $E_{V}$ represents the self-energy of a vortex configuration, and $E_{\text{VV}}$ is the vortex-vortex interaction energy which is non-zero in a vortex (or a anti-vortex) lattice  environment. The expression for the vortex-Skyrmion interaction energy is
\begin{equation}\label{Total Energy}
  E_{\text{int}}=  \int_0^\infty \rho d\rho \left[ -2\pi B_z n^z\right]=  \frac{4\mu}{q \rho_M\lambda^2} \frac{2 \pi R^2 \lambda ^2 \left(\sqrt{R^2+2 \pi  \kappa ^2 \lambda ^2}-\sqrt{2 \pi } \kappa  \lambda  \text{ArcSinh}\left[\frac{\sqrt{2
\pi } \kappa  \lambda }{R}\right]\right)}{\left(R^2+2 \pi  \kappa ^2 \lambda ^2\right)^{3/2}},
\end{equation}
where for the spin configuration we have used the standard Skyrmion configuration represented by $F(u)= \pi - \kappa u$ as an approximation and $\kappa$ is an undetermined constant \cite{Han:2017fyd}. Also, the vortex self-energy is given by 
\begin{eqnarray}
    E_{V}= \frac{1}{\pi \lambda} \left(\frac{\phi_0}{2}\right)^2 \ln{\left(\frac{2 \lambda}{\xi}\right)}.
\end{eqnarray}
Lastly, the vortex-vortex interaction energy is~(see Eq~\eqref{dual action} in Appendix~\ref{Appendix II})
\begin{equation}
    E_{\text{VV}}= \frac{1}{2\pi \lambda^2}\,\,\int_0^\infty \rho d\rho \,\, \vec{\Sigma}\cdot \left(\frac{1}{-\nabla^2 +\tilde{m}^2}\vec{\Sigma}\right).
\end{equation}
{Now from Eq.~\eqref{Total Energy N}, we complete the square to get
\begin{equation}
    E= (E_{V}+ E_{\text{VV}})\left(N + \frac{E_{\text{int}}}{2 (E_{V}+ E_{\text{VV}})}\right)^2 - \frac{E^2_{\text{int}}}{4 (E_{V}+ E_{\text{VV}})}\,.
\end{equation}
%
For $2\lambda > \xi$\, as expected for a type-II superconductor, $E_V>0$\,, and for a vortex (or a anti-vortex) lattice environment the interaction energy $E_{VV}>0$ as well. 
Then the energy is minimum when
\begin{eqnarray}\label{minimum N}
    N= -\frac{E_{\text{int}}}{2 (E_{V}+ E_{\text{VV}})}\,.
\end{eqnarray}
 Thus the winding number of the vortex will be the integer closest to the quantity on the right hand side of Eq.~\eqref{minimum N}.
This can depend on the other parameters of the system, such as $R,\kappa,\lambda,\mu$. We note that for a Skyrmion (anti-Skyrmion) configuration, for which $F(u)= \pi - \kappa u$ (resp. $F(u)= \kappa u$), the sign of vorticity $N$ is negative (resp. positive). This coupled behavior is in agreement with our previous analysis.} {For $E_{\text{int}}> 0\,,$ the parameters $q,\, R,\, \lambda,\, \kappa\,$ must satisfy the condition
%
\begin{eqnarray}
    \frac{1}{q}\left[\sqrt{1+ \left(\frac{\sqrt{2
\pi } \kappa  \lambda }{R}\right)^2}-\frac{\sqrt{2
\pi } \kappa  \lambda }{R}  \text{ArcSinh}\left[\frac{\sqrt{2
\pi } \kappa  \lambda }{R}\right]\right] > 0\,.
\end{eqnarray}
}
\begin{figure}
    \centering
    \includegraphics[width=8cm, height=6cm]{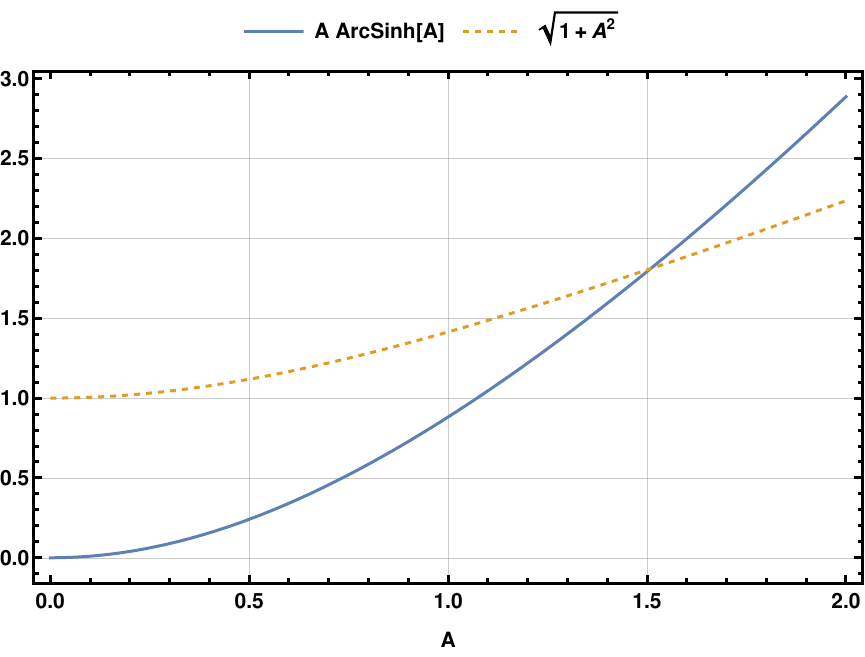}
    \caption{Critical value for the dimensionless constant $A\equiv\dfrac{\sqrt{2\pi } \kappa  \lambda }{R}$}
    \label{fig:9}
\end{figure}
We can plot the two expressions in the square brackets to obtain the critical value for the dimensionless constant $A=\dfrac{\sqrt{2\pi } \kappa  \lambda }{R}$ for satisfying the above condition. As we can see from Fig.~\ref{fig:9}, the value of this constant is $A\simeq1.5$. Thus below this critical value the composite will form. 
\section{Spin waves in {Ginzburg-Landau + nonlinear sigma model} }\label{section-IV}
 In the previous sections we have shown that in our model a Skyrmion like spin configuration arises within a length scale determined by penetration depth and the ratio of coupling constants, as well as the winding number of vortex configuration. However, this is a static configuration of the unit spin field $\mathbf{\hat{n}}$. In general, the spin configurations in such systems can be time dependent. As mentioned in the introduction, the study of spin waves in such systems provides a possible way to probe their magnetic features.  In this section we shall consider the spin field to be time dependent and try to derive an exact propagating wave solution from the Euler-Lagrange equation, both in the absence and presence of vortices. We start from the action for the NLS+GL model with time dependent spin field $\mathbf{\hat{n}}(t)$\,,
\begin{equation}\label{nls-sup-time}
\mathcal{L}= -\frac{1}{4} F_{\mu\nu}^2 + M_0\vec{C}(\mathbf{\hat{n}})\cdot \partial_t \mathbf{\hat{n}} - \frac{\rho_M}{2}\left(\mathbf{\grad} n^i\right)^2 + \frac{\rho_s}{2m}(\partial_\mu\theta - qA_\mu)^2 -\mu n^i (\varepsilon^{ijk} \partial^j A^k + \delta^{iz} B_0) +\rho_M \, Q \,\varepsilon^{ijk} n^i \partial^j n^k\,,
\end{equation}
where we have written the DMI coefficient $D= \rho_M Q$ and $\vec{C}(\mathbf{\hat{n}})$ is introduced as the conjugate momentum for the field 
$\mathbf{\hat{n}}(t)$\,. $M_0$ is the spin per unit volume as defined earlier, $M_0= S\,a^{-3}$, and $\vec{B}_0 \equiv B_0\hat{{z}}$ is the effective background magnetic field due to spontaneous magnetization existing in ferromagnet. The equation of motion of the field $\mathbf{\hat{n}}$ from the above action is given by
\begin{equation}\label{berry con.}
M_0\left(\partial_{n_j} C_i - \partial_{n_i} C_j\right)\partial_t n_i + \rho_M \nabla^2 n_j + 2 \rho_M \, Q \,\varepsilon_{jkl}\partial_k n_l + \mu (B_j + \delta_{jz} B_{0j})=0.
\end{equation}
Here we have denoted by $\partial_{n_i}$ a partial derivative with respect to $n_i$. 
{ The conjugate vector field $\vec{C}$ obeys \cite{PhysRevLett.75.3509, 2010arXiv1009.1603N}
\begin{equation}\label{Berry con.2}
\left(\partial_{n_i} C_j - \partial_{n_j} C_i\right)= \varepsilon_{ijk} n_k\,,
\end{equation}
using which we get the Landau-Lifshitz equation 
\begin{equation}\label{spin wave1}
M_0\partial_t \mathbf{\hat{n}}= \rho_M\, \mathbf{\hat{n}}\times \nabla^2 \mathbf{\hat{n}} + 2\rho_M \, Q\, \left(\mathbf{\hat{n}}\cdot \mathbf{\grad}\right)\mathbf{\hat{n}} + \mu\,\mathbf{\hat{n}}\times (\vec{B}+\vec{B}_0)\,.
\end{equation}
}
To derive the spin wave solution we shall solve this equation along with the wave equation for magnetic field
 \begin{equation}\label{spin wave2}
\left(\partial^2_t-\nabla^2+ \frac{q^2\rho_s}{m}\right)\vec{B}= \frac{q\rho_s}{m}\mathbf{\grad}\times(\mathbf{\grad}\theta) + \mu \mathbf{\grad}\times(\mathbf{\grad}\times\mathbf{\hat{n}}). 
 \end{equation}
 %
 \subsection{Spin wave dispersion in absence of vortex}
We shall first try to solve for the case where no vortex is present i.e. $\mathbf{\grad}\times(\mathbf{\grad}\theta)=0$. For the spin waves propagating along $\mathbf{\hat{z}}$ we take the following ansatz~~\footnote[4]{F.~S.~Nogueira, private communication.}
\begin{equation}\label{spin wave ansz.}
\mathbf{\hat{n}}= \sin\alpha \left[\cos\left((k+Q)z - \omega t\right) \mathbf{\hat{x}}+ \sin\left((k+Q)z - \omega t\right) \mathbf{\hat{y}}\right]+ \cos\alpha \mathbf{\hat{z}}.
\end{equation}
Here we have taken the wave to be circularly polarized and have chosen the polarization to be right handed one. We shall continue with this notation through all subsequent sections. We can find the expression of the magnetic field for this spin wave ansatz using the method of Green functions,
\begin{equation}
   B^i= \int d^4x^\prime \,\,G_W(x-x^\prime) \left[-\mu \nabla^2 n^i\right](x^\prime)\,.
   \label{B.spin-wave}
\end{equation}

Using Fourier transform, we can write the Green function as
\begin{align}
{G_W({x}-{x}^\prime)= \int \frac{d^4p}{(2\pi)^4} \frac{1}{-p_0^2 + p^2 + \tilde{m}^2 }  e^{i(p_0(t-t^\prime) - \vec{p}\cdot(\vec{x}-\vec{x^\prime}))}\,,}
\end{align}
where we have written $\tilde{m} = \dfrac{q^2\rho_s}{m}\,$ as before.
We get an expression for $B^x$\, from this,
\begin{equation}
  B^x=   \frac{\mu \sin\alpha (k+Q)^2}{-\omega^2 + (k+Q)^2 + \tilde{m}^2 } \cos{(-\omega t + (k+Q)z)}\,,
\end{equation}
and an expression for $B^y$ in a similar manner,
\begin{equation}
    B^y=   \frac{\mu \sin\alpha (k+Q)^2}{-\omega^2 + (k+Q)^2 + \tilde{m}^2 } \sin{(-\omega t + (k+Q)z)}\,.
\end{equation}
However, as $\nabla^2 n^z =0$\,, there is no $z$ component of $\vec{B}$. Therefore the total solution can be written as
\begin{equation}\label{T.sol. B}
    \vec{B}= \frac{\mu  (k+Q)^2}{-\omega^2 + (k+Q)^2 + \tilde{m}^2 } \mathbf{\hat{n}} - \frac{\mu \cos\alpha (k+Q)^2}{-\omega^2 + (k+Q)^2 + \tilde{m}^2 } \mathbf{\hat{z}}\,.
\end{equation}

With this expression in hand, one can now put the ansatz of Eq.~\eqref{spin wave ansz.} for the spin field  into Eq.~\eqref{spin wave1} and find the following equation involving the frequency $\omega$ and wave vector $k$\,,
\begin{equation}\label{disp.rel.}
   \omega -\omega_0 - a\left(k^2-Q^2\right)  +\frac{b (k+Q)^2}{-\omega^2 + (k+Q)^2 + \tilde{m}^2 }  =0\,,
\end{equation}
where we have written $a=\dfrac{\rho_M \cos\alpha}{M_0},\,\, b= \dfrac{\mu^2 \cos\alpha}{M_0},\,\,\omega_0= \dfrac{\mu B_0}{M_0}$. Clearly, the last term of  Eq.~\eqref{disp.rel.} is the contribution from the dynamically generated magnetic field and therefore is in general dependent on $\omega, k$. Thus in absence of this contribution we recover the usual dispersion relation for spin waves in a ferromagnet: $ \omega = \omega_0 + a\left(k^2-Q^2\right)$.

\begin{figure}[H]
    \centering
    \subfloat[$\dfrac{b}{a} < \Tilde{m}^2$\,(Green),\,\,$\dfrac{b}{a} > \Tilde{m}^2$\,(Blue),\, $\omega_0=1$ ]{\includegraphics[width=0.35\textwidth]{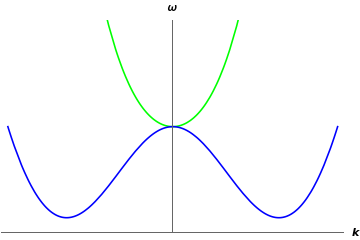}\label{fig:10a}}
    \hspace{0.2cm}
    \subfloat[$\dfrac{b}{a} < \Tilde{m}^2$\,(Cyan),\,\,$\dfrac{b}{a} > \Tilde{m}^2$\,(Purple),\, $\omega_0=0.4$ ]{\includegraphics[width=0.35\textwidth]{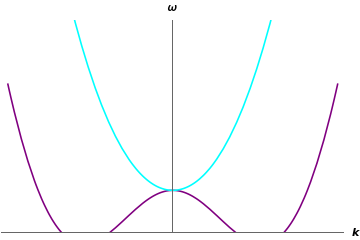}\label{fig:10b}}
   \caption{{Dispersion of spin waves at different values of $a$, $b$, $Q$, and $\Tilde{m}$\,, for $\omega \ll \Tilde{m}$ }}
\end{figure}

\begin{figure}[H]
    \centering
     \subfloat[$Q>0$,\,\, $\omega_0=1$ (Red),\,\, $\omega_0=0.4$ (Magenta) ]{\includegraphics[width=0.35\textwidth]{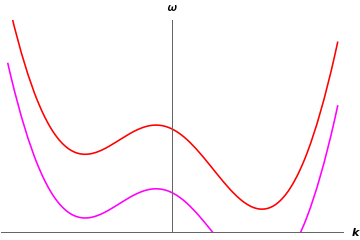}\label{fig:11a}}
     \hspace{0.1cm}
    \subfloat[$Q<0$,\,\, $\omega_0=1$ (Blue),\,\, $\omega_0=0.4$ (Cyan)]{\includegraphics[width=0.35\textwidth]{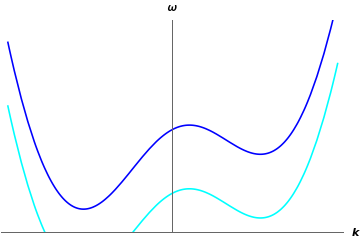}\label{fig:11b}}
    \caption{Dispersion of spin waves at different values of $a$, $b$, $Q$, and $\Tilde{m}$\,, for $\omega \ll \Tilde{m}$}
    \label{fig:11}
\end{figure}

\begin{figure}[H]
    \centering
    \subfloat[$Q>0$]{\includegraphics[width=0.3\textwidth]{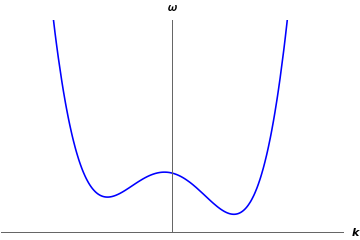}}
    \hspace{0.2cm}
    \subfloat[$Q=0$]{\includegraphics[width=0.3\textwidth]{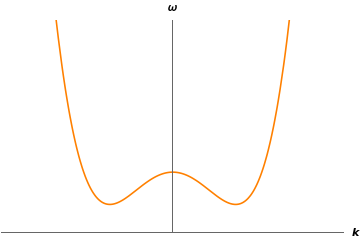}}
    \hspace{0.2cm}
    \subfloat[$Q<0$]{\includegraphics[width=0.3\textwidth]{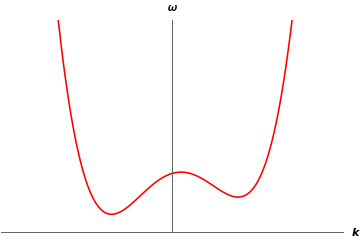}}
    \caption{Dispersion of spin waves at different values of $a$, $b$, $Q$, $\Tilde{m}$ for $\omega \sim \Tilde{m}$.}
    \label{fig:12}
\end{figure}

To understand the physical meaning of the above cubic equation we shall try to derive approximate solutions corresponding to different regions of the frequency of spin waves. For that we shall compare the frequency of the spin waves with the characteristic scale of the superconductor i.e. the photon mass $\Tilde{m}$.  
First we consider the low frequency limit in which the frequency is much lower than photon mass, $\omega^2\ll \tilde{m}^2$. In this approximation we get 
{\begin{equation}\label{approx1}
\begin{aligned}
    \omega- \omega_0 &= a\left(k^2-Q^2\right)  -\frac{ b (k+Q)^2}{  (k+Q)^2 + \tilde{m}^2 }\left[1- \frac{\omega^2}{(k+Q)^2 + \tilde{m}^2}\right]^{-1}
      \\
    & \simeq   a\left(k^2-Q^2\right)  -\frac{b (k+Q)^2}{  (k+Q)^2 + \tilde{m}^2 } - \frac{b (k+Q)^2}{  [(k+Q)^2 + \tilde{m}^2]^2 } \omega^2.    
    \end{aligned}
\end{equation}
The last line of the above equation lead to the following simple quadratic equation
    \begin{align}\label{quadratic_bs}
      &  \omega^2 + \omega_c\, \omega - \omega_c (\omega_0 + \omega_{\text{BS}}) =0,\\
\mathrm{where}\qquad      & \omega_{\text{BS}}= a\left(k^2-Q^2\right)  -\frac{b (k+Q)^2}{  (k+Q)^2 + \tilde{m}^2 },\quad
     \omega_c= \dfrac{  [(k+Q)^2 + \tilde{m}^2]^2 }{b (k+Q)^2}.
    \end{align}
The solution of Eq.~\eqref{quadratic_bs}  is
    \begin{equation}\label{BS_quadratic_freq}
        \omega= -\frac{\omega_c}{2} \pm \frac{\omega_c}{2} \sqrt{1 +4 \frac{(\omega_0 + \omega_{\text{BS}})}{\omega_c}}.
    \end{equation}
Then in the mentioned approximation, which implies $\omega_c \gg \omega_0, \omega_{BS}$, we find that 
    \begin{equation}
        \omega\simeq -\frac{\omega_c}{2} \pm \frac{\omega_c}{2} [1 +2 \frac{(\omega_0 + \omega_{\text{BS}})}{\omega_c}].
    \end{equation}
Considering only the positive sign we get 
\begin{equation} 
\omega= \omega_0 + \omega_{\text{BS}}= \omega_0 + a\left(k^2-Q^2\right)  -\frac{b (k+Q)^2}{  (k+Q)^2 + \tilde{m}^2 },
\end{equation}
which is identical to the result obtained by Braude and Sonin~\cite{PhysRevLett.93.117001} with  $Q=0,\, \cos\alpha>0$.} This dispersion relation has a particular feature that the minimum in the spin wave dispersion curve appears at some finite value of momentum $k=k_{Q=0}=k_0$. Therefore around that point two different spin wave modes exist: one with positive group velocity and another with negative group velocity. This is a distinctive feature for the ferromagnetic superconductors. In presence of a DMI term with a small value of $Q$ this minimum changes. If we denote the $k$ value at which the minimum occurs for a DMI coefficient $Q$ as $k_Q$, it is easy to see that $k_Q$ satisfies the equation
\begin{equation}
    k_Q\left[(k_Q+Q)^2 + \Tilde{m}^2\right]^2= \frac{\mu^2}{\rho_M}\left(k_Q + Q\right)\Tilde{m}^2.
\end{equation}
From the above equation we can see that at $Q=0$ we have three real solutions : $k_0=0$ and $k_0= \pm \Tilde{m}\sqrt{\dfrac{1}{\Tilde{m} } \sqrt{\dfrac{\mu^2}{\rho_M}} -1 }$\,. It was shown in~\cite{PhysRevLett.93.117001} that for  $\sqrt{\dfrac{\mu^2}{\rho_M}} < \Tilde{m}$ there is a minimum at $k=0$, while for $\sqrt{\dfrac{\mu^2}{\rho_M}} > \Tilde{m}$ the minimum shifts to $k=k_0$.
This is depicted in Fig.~\ref{fig:10a} and Fig.\ref{fig:10b} for two different values of $\omega_0$ and $\tilde{m}=10$, $Q=0$. {The wave stops propagating when the frequency goes to zero, so that for a sufficiently small $\omega_0$\,, the spin wave disappears for some values of $k$, leading to a break in the spectrum. }
The spin waves with $k< k_0$ will propagate with a negative group velocity while for the spin waves with $k> k_0$ the group velocity 
is positive. In presence of a DMI term one of the two minima has lower energy, depending on the sign of $Q$, 
as shown in Fig.~\ref{fig:11a} and \ref{fig:11b} where we have taken $\Tilde{m}=10$ eV, $a=0.06$ eV$^{-1}$, $b=10$ eV, $|Q|=0.4$ eV. 
In the small $Q$ approximation the location of the two minimum points of the dispersion curve can be written as
\begin{equation}
    k_Q \approx \pm (k_0 - Q)+\cdots
\end{equation}
Thus we can see that in presence of DMI term the location of $k_0$ gets shifted by $Q$. A similar analysis can be done for $\cos\alpha<0$. In this case, depending on whether $\dfrac{b}{a}$ is smaller or larger than $\Tilde{m}^2$ there can be one or two maxima. 
Next let us consider the region where the frequency is close to the photon mass, i.e. $|\omega- \Tilde{m}|$ is small compared to $\omega$ or $\tilde{m}$. In this case one can expand the cubic equation as
\begin{equation}\label{approx2}
\begin{aligned}
    &\omega-\omega_0 = a\left(k^2-Q^2\right)  -\frac{b (k+Q)^2}{-\omega^2 + (k+Q)^2 + \tilde{m}^2 }  \\
    &= a\left(k^2-Q^2\right)  -b \left[1- \frac{\omega^2-\Tilde{m}^2}{(k+Q)^2 }\right]^{-1}
      \\
    & \simeq  a\left(k^2-Q^2\right)  - b \left[1+ \frac{\omega^2-\Tilde{m}^2}{(k+Q)^2 }\right]+ \cdots \,.
    \end{aligned}
\end{equation}
 The solution of this quadratic equation is 
\begin{eqnarray}
    \omega= -\frac{(k+Q)^2}{2b} \pm \sqrt{\Tilde{m}^2 +\frac{(k+Q)^4}{b^2}- (k+Q)^2 \left[1-\frac{\omega_0}{b}-\frac{a}{b}(k^2-Q^2)\right]}\,.
\end{eqnarray}
%
In Fig.~\ref{fig:12} we have displayed a particularly interesting case of this solution, 
which describes the dispersion in presence of superconductivity, {for the choices} $\Tilde{m}=4$, $\dfrac{b}{a}\sim 1$ and with the DMI coefficient $Q$ taken to be {$Q=-0.1,\, 0,\, 0.1$}. At $Q=0$ the dispersion curve describes two degenerate minima at some non-zero value of wave vector $k$. As we start to increase $Q$ from $0$ the degeneracy breaks and two non-degenerate minima develop on the dispersion curve.
The excitation around the local minimum with a finite gap describes a roton like mode. Such roton like character in spin waves was proposed theoretically for different magnetic systems and termed magnetic rotons~\cite{PhysRevB.99.014407, PhysRevLett.93.036405}. Thus we can say that the spin wave excitations near to these minima shown in Fig.~\ref{fig:11} and Fig.~\ref{fig:12} describes magneto roton excitations which is a very interesting feature.

Lastly, one can take the frequency of the spin waves to be higher than the photon mass i.e. $\omega^2 \gg \Tilde{m}^2$. In this limit, the effect of superconductivity on the the spin wave spectrum will be suppressed and the nature of spin waves will be similar to that in a ferromagnet coupled to electromagnetic fields. The equation governing the behavior of spin waves in this regime is given below
\begin{equation}\label{disp.rel.1}
    \omega -\omega_0- a\left(k^2-Q^2\right)  +\frac{b (k+Q)^2}{-\omega^2 + (k+Q)^2 }  =0.
\end{equation}
The nature of the spin waves in a ferromagnet with the {SMFI} interaction is a well studied subject and therefore we do not discuss that here. 

\subsection{Effect of a static vortex on spin wave solution}
Let us now introduce vortices in the theory. These act as a source of an inhomogeneous magnetic field. Due to this inhomogeneous field the amplitude of the spin field cannot remain constant through out the system. In particular, in the presence of a cylindrically symmetric Abrikosov vortex along the $z$ axis, a radially varying magnetic field profile is created, decaying sharply with distance. In such a case, let us take the following ansatz for the spin wave inspired from that in Eq.~\eqref{spin wave ansz.}
\begin{equation}\label{ansatz-2}
   M_0\, \mathbf{\hat{n}}= \sin\alpha(\rho) \left[\cos\left((k+Q)z - \omega t\right) \mathbf{\hat{x}}+ \sin\left((k+Q)z - \omega t\right) \mathbf{\hat{y}}\right]+ \cos\alpha(\rho) \mathbf{\hat{z}}\,,
\end{equation}
where $\rho$ is the radial distance from the axis, while the frequency and wave vector are assumed to be constant. The profile of the function $\alpha(\rho)$ has to be determined from the Landau-Lifshitz equation of motion for $\mathbf{\hat{n}}$. For this let us again write down again the Landau-Lifshitz equation of Eq.~\eqref{spin wave1}
\begin{equation}\label{Landau-Lifshitz}
M_0\,\partial_t \mathbf{\hat{n}}= \rho_M\, \mathbf{\hat{n}}\times \nabla^2 \mathbf{\hat{n}} + 2\rho_M \, Q\, \left(\mathbf{\hat{n}}\cdot \mathbf{\grad}\right)\mathbf{\hat{n}} + \mu\, \mathbf{\hat{n}}\times \vec{B}\,.
\end{equation}

As the DMI term will bring a dependence on the angular coordinate $\phi$, in presence of this term solution will be much harder to derive. For simplicity, we set $Q=0$ and proceed to solve the equation. We showed earlier that the magnetic field due to a cylindrically symmetric Abrikosov vortex has the profile
\begin{align}
{B^z= \frac{1}{2\pi \lambda^2 q}\int d^3x^\prime \,\,  G(\vec{x}-\vec{x}^\prime)  \Sigma^z(\vec{x}^\prime)= \frac{2N}{\lambda^2 q}K_0\left(\frac{\rho}{\sqrt{2\pi}\lambda}\right)\,,}
\end{align}
where $G(\vec{x}-\vec{x}^\prime)$ was defined in Eq.~(\ref{3d.Green}).
Adding this magnetic field of a vortex to the field for the spin wave as found in Eq.~(\ref{B.spin-wave})\,, we can rewrite 
Eq.~\eqref{Landau-Lifshitz} as
\begin{align}\label{LL with Vortex}
    {M_0\,\partial_t\mathbf{\hat{n}}=  \rho_M \mathbf{\hat{n}}\times \,\nabla^2\mathbf{\hat{n}}+ \frac{2\mu N}{\lambda^2 q}K_0\left(\frac{\rho}{\sqrt{2\pi}\lambda}\right)\,\mathbf{\hat{n}}\times \mathbf{\hat{z}} - \mu^2 \mathbf{\hat{n}}\times\int d^4x^\prime 
     G_W({x}-{x}^\prime)\, \nabla^2\mathbf{\hat{n}}({x}^\prime)\,.}
\end{align}
Putting the ansatz of Eq.~\eqref{ansatz-2} into Eq.~\eqref{LL with Vortex} we obtain  for the field $\alpha(\rho)$
\begin{equation}\label{spin wave w. vortex2}
    \partial^2_\rho \alpha +\frac{1}{\rho} \partial_\rho\alpha - \frac{k^2}{2}\sin(2\alpha(\rho)) - \left(\frac{2\mu N}{\lambda^2 q \rho_M} K_0\left(\frac{\rho}{\sqrt{2\pi}\lambda}\right)-\frac{M_0\,\omega}{\rho_M}\right) \sin\alpha +\frac{\mu^2}{\rho_M}  \zeta\left(\omega, k, \rho\right) =0\,,
\end{equation}
where we have written
\begin{equation}
\begin{aligned}
& \zeta\left(\omega, k, \rho\right)=\int d^2x_{\parallel}^\prime d^2p_\parallel\,\, e^{-i\textbf{p}_\parallel\cdot \left(\textbf{x}-\textbf{x}^\prime\right)_\parallel} \left[\cos\alpha(\rho) f_1(\rho^\prime, \omega, k, \Tilde{m}) + \sin\alpha(\rho) f_2(\rho,\Tilde{m})\right]\,,\\ &
 f_1(\rho^\prime, \omega, k, \Tilde{m})= \frac{\left[\cos\alpha(\rho^\prime)\left(\partial^2_{\rho^\prime} \alpha(\rho^\prime) +\frac{1}{\rho^\prime} \partial_{\rho^\prime}\alpha(\rho^\prime)\right) - \sin\alpha(\rho^\prime)(\partial_{\rho^\prime}\alpha(\rho^\prime))^2- \frac{k^2}{2}\sin(\alpha(\rho^\prime))\right]}{\textbf{p}^2_{\parallel} + k^2 +\Tilde{m}^2 - \omega^2}\,,\\ &
 f_2(\rho^\prime, \Tilde{m})= \frac{\left[\sin\alpha(\rho^\prime)\left(\partial^2_{\rho^\prime} \alpha(\rho^\prime) +\frac{1}{\rho^\prime} \partial_{\rho^\prime}\alpha(\rho^\prime)\right)+\cos\alpha(\rho^\prime)(\partial_{\rho^\prime}\alpha(\rho^\prime))^2 \right]}{\textbf{p}^2_{\parallel} +\Tilde{m}^2\,.}
\end{aligned}
\end{equation}

Here the symbols $x_{\parallel},\,\, p_{\parallel}$ denote the $x,\, y$ components of the position
and momentum vectors. The field $\zeta(\omega, k, \rho)$ is the magnetic field generated by the spin field $\mathbf{\hat{n}}$ and has to be determined in a self-consistent way. Eq.~\eqref{spin wave w. vortex2} governs the behavior of the field $\alpha(\rho)$ for a given constant frequency $\omega$ and wave number $k$. 
As we can see, this equation resembles Eq.~\eqref{Effect of vortex on Skyr.} where we had not taken $\mu^2$ term for simplicity. Thus the amplitude of the spin wave give rise to a spatial profile similar to the Skyrmion solution shown previously. Solving Eq.~(\ref{spin wave w. vortex2}) should produce a more detailed understanding of the behavior of spin waves in the presence of a vortex. We leave that for later work.

\section{Discussion and Outlook}\label{section-V}
In this work we have considered a nonlinear sigma model coupled to a Ginzburg Landau model of superconductivity in the Landau limit via an SMFI coupling term. Such a model would describe a ferromagnetic superconductor and topological excitation that may appear in such systems. Here we have analyzed how topological structures, like vortices in superconducting order and Skyrmions in the ferromagnetic order, can become intertwined. We have shown that for a straight vortex along the $z$-axis, a Skyrmion like configuration arises within a finite region with a size depending on the London penetration depth of the superconductor and also on other parameters of the model.\\

This indicates that in a system described by the coupled models, a coupled structure of Skyrmion and vortex does indeed appear. In addition, we have shown that this particular solution arises only when the spin at the origin aligns along the magnetic field of the vortex. This suggests that such systems, if stabilized, can manifest a coupled motion of the Skyrmion and vortices. It has been known for a long time that in the presence of an external current $\textbf{J}_T$\,, a moving vortex with velocity $\textbf{V}_L$ experiences a force transverse to the direction of its velocity, given by~\cite{1990PZETF..52.1141F}
\begin{equation}
    F_L= \pi \hbar (n_\infty - n_0) \hat{\textbf{z}}\times \textbf{V}_L,
\end{equation}
where $n_0$ and $n_\infty$ are the particle densities at the core and far outside the vortex, respectively. This force is independent of the charge and not of electromagnetic origin. For a vortex coupled to a Skyrmion in the manner described in this paper this transverse force will produce a motion of the vortex which in turn will produce a motion of the Skyrmion perpendicular to the applied current. The motion of a Skyrmion perpendicular to the applied current is famously known as Skyrmion Hall effect~\cite{2008PhR...468..213T,nnano.2013.243} which was recently discovered~\cite{2017NatPh..13..162J}. In that context the Skyrmion Hall effect appears due to the torque applied by the conduction electron spin via a $s-d$ type exchange interaction of the local spins and conduction electron spin~\cite{2008PhR...468..213T}. However, in the present case the Skyrmion Hall effect, if found, would be a manifestation of the coupling between the vortex and Skyrmion like structures.\\

{It was previously shown that ferromagnetic superconductors in two and three dimensions can act as a host to finite density of zero energy Majorana surface modes~\cite{PhysRevB.85.144505}. In comparison to such modes, the Majorana modes at the vortex core or Skyrmion centre seem more advantageous for quantum computing applications due to their localized nature and possibility of braiding through vortex motion~\cite{RevModPhys.87.137}. Since the Skyrmion-vortex composite described in this work may also host such Majorana fermions, it opens up possibilities of  further work evaluating the potential of the bulk ferromagnetic-superconductors for detecting Majorana fermions.}\\

We note that while solving the equations for the function $F(\rho)$ we have always considered cylindrical symmetry. For this, the solutions will be valid for any plane perpendicular to the object's axis which in our case is the $z$-axis. Thus our solution describes a cylindrically symmetric three dimensional structure having a finite energy per unit length along the axis -- its total energy increases with its length along the axis. For a thick bulk sample the energy of such a configuration may get out of the realizable limit in condensed matter systems. To avoid this problem one can focus on ferromagnetic superconductors made as thin films. The small thickness of the sample would thus allow such a composite excitation to extend only over a few nanometers of distance along the axis and thus its energy will be within the experimentally realizable limit.\\

We also note that the interaction between superconductivity and ferromagnetism considered in this paper occurs only via the magnetic fields generated by static spins. One can ask how the Skyrmion-vortex composite configuration would modify if one include direct interaction terms such as those considered by Greenside et al~\cite{PhysRevLett.46.49}, Ng et al~ \cite{PhysRevB.58.11624}. To answer this let us first write down the direct coupling terms considered by these authors,
\begin{equation}
    E \sim \eta_1 |\vec{M}|^2 |\psi|^2 + \eta_2 |\partial_i\vec{M}|^2 |\psi|^2\, \qquad \eta_1\sim \frac{T_{\text{mag}}}{T_C}\, \qquad \eta_2\sim \xi^2 \eta_1\,,
\end{equation}
where $T_{\text{mag}}$ and $T_C$ are the ferromagnetic and superconducting transition temperature respectively and $\xi$ is the coherence length of the superconductor. It was argued in~\cite{PhysRevLett.46.49} that, as  $T_{\text{mag}} \ll T_C$ for the rare earth compounds that we have mentioned in the introduction, the direct coupling terms will not be relevant. Although this argument is still valid for thin film samples of rare earth compounds, it fails for other systems mentioned by us like Hydrazine treated NbSe$_2$~\cite{Zhu2016SignatureOC}. However, we note that we have taken the approximation that the amplitude of the condensate is a constant outside the vortex and also {that} $\vec{M}= \mu\, \mathbf{\hat{n}}$ --- then the first term of the expression of $E$ above is a constant, while the second term of $E$ renormalizes the second term of the action of Eq.~\eqref{nls-sup}. Thus with these approximation the direct coupling term would not affect the nature of our action and hence the solutions for the Skyrmion-vortex composite remain unchanged.\\

The spin wave solution in such a model also shows some very interesting features as we have discussed. In particular roton like gapped excitation appear around the local minimum in dispersion curve. The dispersion of spin waves in these gapped local minima resembles that of a massive particle. This suggests the intriguing possibility of an Anderson-Higgs mass generation mechanism for magnons which was recently discussed in a superconductor-ferromagnet-superconductor heterostructure~\cite{PhysRevApplied.18.L061004}. Constructing a field theoretic description of the Anderson-Higgs mass generation for magnons, starting from a microscopic theory of the present system and using effective bosonic representations, is an interesting possibility for future work.\\ 

We mentioned that as soon as frequency of the spin waves approaches zero it becomes non-propagating, leading to a discontinuity in the dispersion curves of Fig.~\ref{fig:10b} and Fig.~\ref{fig:11}. With the approximations we had taken in Eq.~\eqref{approx1}, the $\omega-k$ curve crosses the $\omega=0$ axis at most twice for a given direction of propagation. However, one can consider other higher order terms which would lead to multiple such crossings. This would create a band like formation in $\omega-k$ diagram.\\

The spin wave dispersion i.e. the $\omega-k$ relation, comes from solving Eq.~\eqref{disp.rel.}, a cubic equation, in the absence of a vortex --- a more general equation needs to be solved in the presence of a vortex. In general, this  differs from the quadratic dispersion which is usually seen in magnetic systems. A change in the $\omega-k$ relation can lead to a modification of the thermodynamic property of a collection of magnonic excitations. We know that magnons influence the electronic specific heat as well as thermal and electrical conductivity of magnetic systems~\cite{PhysRevLett.17.750}. Thus a change in nature of spin wave dispersion affect these properties which would be experimentally observable. This also opens a possible direction for future investigation. 

\section*{Acknowledgment}

\noindent {This work was partially carried out during a research visit by SM to the Leibniz Institute supported by an MoU between the S. N. Bose Centre and the Leibniz Institute.}
{The authors gratefully acknowledge many discussions with Flavio Nogueira who provided many insightful ideas which proved to be essential for the progress and completion of this project. SM also acknowledges discussions with Arijit Haldar, Shibendu G. Choudhury, and Ramesh Pramanik. }

\appendix

\section{Values of different constants for $\text{NbSe}_2$ ($\hbar=c=1$) }\label{Appendix I}
\begin{itemize}
    
\item \textbf{Mean field method of determining the value of the exchange coupling J}\\

If the ferromagnetic transition temperature is $T_c$ then 
\begin{equation}
    k_B T_c \sim  \frac{2}{3} Z J S(S+1),
\end{equation}
where $Z$ is the number of nearest neighbors, $S$ is the spin per site,  $k_B$ is the Boltzmann constant. For NbSe$_2$ the transition temperature is $T_c \sim 40$K and $Z=6$\cite{Zhu2016SignatureOC}. {These values give}
\begin{equation}
    J = \frac{2 k_B T_c }{z }= 1.146\,{\rm meV}
\end{equation}
\item \textbf{Determination of exchange stiffness $\rho_M$}\\

We use the formula $\rho_M\sim \frac{J S^2}{a}$, where $a$ is the lattice constant. For $\text{NbSe}_2$\,, we have $a= 0.344$\,nm. Then we can estimate $\rho_M$ in eV units as 
\begin{eqnarray}
    \rho_M\sim \frac{J S^2}{a} \sim 0.165 \,\text{eV}^2.
\end{eqnarray}

\item \textbf{Determination of Magnetization $\mu$}\\

In the region of coexistence of superconductivity and ferromagnetism i.e. below $\sim 7 K$, the saturation magnetization in NbSe$_2$ is 
$ 2\,\, \text{emu/g}$ \cite{Zhu2016SignatureOC}. The density of such materials is approximately $6.3 \,\text{g/cm}^3$. Thus the magnetization density in eV units is given by
\begin{equation}
   \mu = 2 \times 6.3\,\text{emu/cm}^3 \sim 3.18\,\, \text{eV}^2\,.
\end{equation}
\item Using all these values we can now determine, for given values of $N$ and $\lambda$\,,
\begin{equation}
    \frac{2N \mu}{q \rho_M \lambda^2}\sim \frac{N}{\lambda^2} 63
\end{equation}
\item \textbf{Penetration Depth }\\
The penetration depth for NbSe$_2$ in its superconducting phase is $\sim 124$ nm~\cite{FINLEY1980493}. However, this is obtained in absence of any magnetic property. After the Hydrazine treatment the substance becomes ferromagnetic and in presence of magnetism the penetration depth must increase to higher values. We have therefore assumed $\lambda$ to be in the range $200$ nm to $1000$ nm, which in eV units becomes 1~eV$^{-1}$ to 5~eV$^{-1}$\,.
\item We take the value of DMI coefficient $D$ to be $\sim 0.1~\text{mJ/m}^2$ which corresponds to $2.496~\text{eV}^3$.
\item Lastly, we note that magnetic field is converted to eV$^2$ unit as $1~T \sim 200~\text{eV}^2$.
\end{itemize}

\section{ Alternative derivation of Eq.~(\ref{Effect of vortex on Skyr.}) via dual formulation }\label{Appendix II}
Here we provide a derivation of Eq.~(\ref{Effect of vortex on Skyr.}) from a dual description of the theory.
Let us write down the generating functional $\mathcal{Z}$ of our system
\begin{equation}
\begin{aligned}
&\mathcal{Z}= \int \mathcal{D}n^i \mathcal{D}A^i \mathcal{D} \theta \exp\left[i \int d^4x \mathcal{L} \right], \\
& \mathcal{L}= -\frac{1}{4} F_{ij}^2 - \frac{\rho_M}{2}\left(\partial_\mu n^i\right)^2 -\frac{\rho_s}{2m}(\vec{\nabla}\theta - q\vec{A})^2 -\mu n^i \varepsilon^{ijk} \partial^j A^k
\end{aligned}
\end{equation}
To proceed, we linearize the third term in the Lagrangian by Hubbard-Stratonovich transformation,
\begin{equation}
\begin{aligned}
&\mathcal{Z}= \int \mathcal{D}n^i \mathcal{D}A^i \mathcal{D} \theta \mathcal{D} f^i \exp\left[i \int d^4x \mathcal{L} \right], \\
& \mathcal{L}= -\frac{1}{4} F_{ij}^2 - \frac{\rho_M}{2}\left(\partial_\mu n^i\right)^2 -2 \sqrt{\frac{\rho_s}{2m}}f^i (\vec{\nabla}\theta - q\vec{A})^i - f^i f^i -\mu n^i \varepsilon^{ijk} \partial^j A^k\,,
\end{aligned}
\end{equation}
where $f^i$ is a new auxiliary vector field. The phase field $\theta$ is multivalued in presence of a vortex. So we can breakdown $\vec{\nabla}\theta$ into two different parts: $\vec{\nabla}\theta= \vec{\nabla}\theta^r + \vec{\nabla}\theta^s$, where the first part is irrotational while the second part is constituted of multivalued field $\theta^s$ and has non-zero curl. Also, the fields $\theta^r$ and $\theta^s$ behave as independent variables. With this decomposition we can write the above partition function as
\begin{equation}
\begin{aligned}
&\mathcal{Z}= \int \mathcal{D}n^i \mathcal{D}A^i \mathcal{D} \theta^r  \mathcal{D} \theta^s  \mathcal{D} f^i \exp\left[i \int d^4x \mathcal{L} \right], \\
& \mathcal{L}= -\frac{1}{4} F_{ij}^2 - \frac{\rho_M}{2}\left(\partial_\mu n^i\right)^2 -2 \sqrt{\frac{\rho_s}{2m}}f^i \partial^i\theta^r-2 \sqrt{\frac{\rho_s}{2m}}f^i (\vec{\nabla}\theta^s - q\vec{A})^i - f^i f^i -\mu n^i \varepsilon^{ijk} \partial^j A^k\,.
\end{aligned}
\end{equation}

We can now integrate out $\theta^r$ functionally to obtain the delta functional $\delta(\vec{\nabla}\cdot \vec{f})$. This imposes the constraint that the vector field $\vec{f}$ is conserved. This can be satisfied by the ansatz 
\begin{equation}
\vec{f} = \frac{1}{2}\vec{\nabla}\times \vec{C},
\end{equation}
where $\vec{C}$ is a new vector field. This ansatz {allows us to rewrite} the delta functional as $\delta(\vec{f} -\frac{1}{2}\vec{\nabla}\times \vec{C})$\,. We can now functionally integrate over $f^i$ to get the action
\begin{equation}
\begin{aligned}
&\mathcal{Z}= \int \mathcal{D}n^i \mathcal{D}A^i \mathcal{D} \theta^s  \mathcal{D} C^i \exp\left[i \int d^4x \mathcal{L} \right], \\
& \mathcal{L}= -\frac{1}{4} F_{ij}^2 - \frac{\rho_M}{2}\left(\partial_\mu n^i\right)^2 -\sqrt{\frac{\rho_s}{2m}}\left(\vec{\nabla}\times \vec{C}\right)\cdot \vec{\nabla}\theta^s + q\sqrt{\frac{\rho_s}{2m}}\left(\vec{\nabla}\times \vec{C}\right)\cdot\vec{A} - \frac{1}{4}\left(\vec{\nabla}\times \vec{C}\right)^2 -\mu n^i \varepsilon^{ijk} \partial^j A^k,
\end{aligned}
\end{equation}

We can define the vorticity as
\begin{equation}
\vec{\Sigma}= \frac{\Phi_0}{2\pi}\,\,\vec{\nabla}\times (\vec{\nabla}\theta^s)\,.
\end{equation}
With this definition and recalling that $\sqrt{\dfrac{\rho_s}{2m}}= \dfrac{1}{\sqrt{4\pi}\lambda}\,\,\dfrac{\Phi_0}{2\pi}$, where $\lambda$ is the penetration depth, we can rewrite the generating functional
\begin{align}
&\mathcal{Z}= \int \mathcal{D}n^i \mathcal{D}A^i \mathcal{D} \theta^s  \mathcal{D} C^i \exp\left[i \int d^4x \mathcal{L} \right]\,, \\
& \mathcal{L}= -\frac{1}{4} F_{ij}^2 - \frac{\rho_M}{2}\left(\partial_\mu n^i\right)^2 -\frac{1}{\sqrt{4\pi}\lambda} \vec{C}\cdot \vec{\Sigma} + \frac{1}{\sqrt{4\pi}\lambda}\left(\vec{\nabla}\times \vec{C}\right)\cdot\vec{A} - \frac{1}{4}\left(\vec{\nabla}\times \vec{C}\right)^2 -\mu n^i \varepsilon^{ijk} \partial^j A^k.
\end{align}


To obtain a theory written in terms of the vorticity $\vec{\Sigma}$ and unit vector field $\mathbf{\hat{n}}$ we integrate out both the vector fields $A^i$ and $C^i$ and obtain the following partition function~\cite{PhysRevB.58.11624, Mukherjee:2019vmi, Mukherjee:2021ypk}
%
\begin{align}\label{dual action}
&\mathcal{Z}= \int \mathcal{D}n^i \mathcal{D} \theta^s  \exp\left[i \int d^4x \mathcal{L} \right]\,, \\
& \mathcal{L}= - \frac{\rho_M}{2}\left(\partial_\mu n^i\right)^2 -\frac{\mu}{2\pi \lambda^2}\,\, \mathbf{\hat{n}}\cdot \left(\frac{1}{\nabla^2 - \tilde{m}^2} \vec{\Sigma} \right) + \frac{1}{2\pi \lambda^2}\,\,\vec{\Sigma}\cdot \left(\frac{1}{\nabla^2 - \tilde{m}^2}\vec{\Sigma}\right)  + \mu^2 \mathbf{\hat{n}} \cdot \left(\frac{\nabla^2}{\nabla^2 - \tilde{m}^2} \mathbf{\hat{n}}\right)\,,\label{dual-lagrangian}
\end{align}
%
where we have written $\tilde{m}^2= \frac{1}{2\pi \lambda^2}$. This is the dual theory we wanted to achieve. We shall now show that the Eq.~\eqref{Effect of vortex on Skyr.} can be easily obtained from the above dual theory. For that we first use the ansatz of Eq.~\eqref{skyr. ansatz} in the Lagrangian of Eq.~(\ref{dual-lagrangian}) to get
\begin{equation}
 \mathcal{L}= \frac{\rho_M}{2} \left(\partial_i F \partial_i F  + \frac{1}{\rho^2}\sin^2 F\right)-\frac{\mu}{2\pi \lambda^2}\,\, \mathbf{\hat{n}}\cdot \left(\frac{1}{\nabla^2 - \tilde{m}^2} \vec{\Sigma} \right),
\end{equation}
where we have ignored the spin-spin interaction term which is a higher derivative term, and also the vortex-vortex interaction term is zero in the scenario where only one vortex is present. The Euler-Lagrange equation for the field $F$ is given by
%
\begin{align}
\nabla^2 F - &\frac{\sin 2F}{2\rho^2} + \frac{\mu}{2\pi \rho_M \lambda^2} \sin F \left(\frac{1}{\nabla^2 - \tilde{m}^2} \Sigma^z \right) \notag \\ &- \frac{\mu}{2\pi \rho_M \lambda^2} \cos F\cos\phi \left(\frac{1}{\nabla^2 - \tilde{m}^2} \Sigma^x\right) - \frac{\mu}{2\pi \rho_M \lambda^2} \cos F\sin\phi \left(\frac{1}{\nabla^2 - \tilde{m}^2} \Sigma^y \right)=0.
\end{align}

For a vortex along the $z$-axis the last two terms vanish and we can write 
\begin{equation}
\nabla^2 F - \frac{\sin 2F}{2\rho^2} - \frac{\mu}{2\pi \rho_M \lambda^2} \sin F \left(\frac{1}{-\nabla^2 +\tilde{m}^2} \Sigma^z \right)=0\,.
\end{equation}
It is easy to see that
\begin{equation}
\left(\frac{1}{-\nabla^2 +\tilde{m}^2} \Sigma^z \right)= N\,\Phi_0\int d^3y\,\, dz^\prime \,\, G(\vec{x}-\vec{y}) \int dz^\prime \delta^3(\vec{y}-\mathbf{\hat{z}} \,\, z^\prime)= 2N\, \Phi_0 \,K_0(\tilde{m}\rho)
\end{equation}
Thus finally we obtain
\begin{equation}
\nabla^2 F - \frac{\sin 2F}{2\rho^2} - N \frac{2\mu\Phi_0}{2\pi \rho_M \lambda^2} \sin F K_0(\tilde{m}\rho)=0\,,
\end{equation}
which agrees with Eq.(\ref{Effect of vortex on Skyr.}), with $\Phi_0 = \dfrac{2\pi}{q}$\,.
\section{Form of the potential in Sec-\ref{section-II}}\label{Appendix III}
It is known that, the equation~\eqref{mod. Skyr.DMI F 2} leads to a Skyrmion configuration only in the presence of a potential term~\cite{2020NatRP...2..492B}, given by~
\begin{equation}
    V(F)= K\left[1 - \mu^2\left(\hat{r}\cdot \mathbf{\hat{n}}\right)^2\right] + \mu \left[B - \mathbf{\hat{n}}\cdot \vec{B}\right].
\end{equation}
The first term describes uniaxial magnetocrystalline anisotropy with the coefficient $K$\,, with $\hat{r}$ being the unit vector in the direction of the high symmetry axis, whereas the second term is the SMFI energy. In the presence of this potential, Eq.~\eqref{mod. Skyr.DMI F 2} will be modified as
\begin{equation}
    \nabla^2 F - \frac{1}{2\rho^2}\sin 2F + \frac{D}{\rho_M \rho}(\cos 2F -1) - K \sin F \cos F - \frac{\mu B}{\rho_M} \sin F=0\,.
\end{equation}
This equation leads to the desired Skyrmion solution.
\medskip
\bibliography{Skyrmion.bib}

\end{document}